\documentclass[a4paper, amsfonts, amssymb, amsmath, reprint, showkeys, nofootinbib, twoside, superscriptaddress, aps]{revtex4-1}

\usepackage{color}
\definecolor{blue-violet}{rgb}{0.54, 0.17, 0.89}

\usepackage{amsthm}
\usepackage[english]{babel}
\usepackage[utf8]{inputenc}
\usepackage[colorinlistoftodos, color=green!40, prependcaption]{todonotes}
\usepackage{mathtools}
\usepackage{physics}
\usepackage{xcolor}
\usepackage{graphicx}
\usepackage[left=23mm,right=13mm,top=35mm,columnsep=15pt]{geometry} 
\usepackage{adjustbox}
\usepackage{placeins}
\usepackage[T1]{fontenc}
\usepackage{lipsum}
\usepackage{csquotes}
\usepackage{tikz}
\usepackage{bm}
\usepackage{braket}
\usepackage{multirow}
\usepackage{qcircuit}
\usepackage{siunitx}

\usepackage{titlesec}
\titlespacing\section{-5pt}{10pt plus 4pt minus 2pt}{2pt plus 2pt minus 2pt}
\titleformat{\section}{\normalfont\bfseries\raggedright}{\thesection}{1em}{}
\titleformat{\subsection}{\normalfont\bfseries\raggedright}{\thesubsection}{1em}{}

\usepackage{hyperref}
\hypersetup{
    colorlinks=true,
    linkcolor=blue,
    urlcolor=blue,
    citecolor=blue
    }
\usepackage[capitalise]{cleveref}

\usepackage{fancyhdr}
\begin{document}

\title{Benchmarking quantum logic operations relative to thresholds for fault tolerance}

\author{Akel Hashim}
    \thanks{Email correspondence: ahashim@berkeley.edu}
    \affiliation{Quantum Nanoelectronics Laboratory, Department of Physics, University of California at Berkeley, Berkeley, CA 94720, USA}
    \affiliation{Graduate Group in Applied Science and Technology, University of California at Berkeley, Berkeley, CA 94720, USA}
    \affiliation{Computational Research Division, Lawrence Berkeley National Lab, Berkeley, CA 94720, USA}
\author{Stefan Seritan}
\author{Timothy Proctor}
\author{Kenneth Rudinger}
    \affiliation{Quantum Performance Laboratory, Sandia National Laboratories, Albuquerque, NM 87185 and Livermore, CA 94550}
\author{Noah Goss}
    \affiliation{Quantum Nanoelectronics Laboratory, Department of Physics, University of California at Berkeley, Berkeley, CA 94720, USA}
\author{Ravi K. Naik}
    \affiliation{Quantum Nanoelectronics Laboratory, Department of Physics, University of California at Berkeley, Berkeley, CA 94720, USA}
    \affiliation{Computational Research Division, Lawrence Berkeley National Lab, Berkeley, CA 94720, USA}
\author{John Mark Kreikebaum}
    \thanks{Now at Google Quantum AI, Mountain View, California, USA.}
    \affiliation{Quantum Nanoelectronics Laboratory, Department of Physics, University of California at Berkeley, Berkeley, CA 94720, USA}
    \affiliation{Materials Sciences Division, Lawrence Berkeley National Lab, Berkeley, California 94720, USA}
\author{David I. Santiago}
    \affiliation{Computational Research Division, Lawrence Berkeley National Lab, Berkeley, CA 94720, USA}
\author{Irfan Siddiqi}
    \affiliation{Quantum Nanoelectronics Laboratory, Department of Physics, University of California at Berkeley, Berkeley, CA 94720, USA}
    \affiliation{Computational Research Division, Lawrence Berkeley National Lab, Berkeley, CA 94720, USA}
    \affiliation{Materials Sciences Division, Lawrence Berkeley National Lab, Berkeley, CA 94720, USA}

\date{\today}

\begin{abstract}
Contemporary methods for benchmarking noisy quantum processors typically measure average error rates or process infidelities. However, thresholds for fault-tolerant quantum error correction are given in terms of worst-case error rates --- defined via the diamond norm --- which can differ from average error rates by orders of magnitude. One method for resolving this discrepancy is to randomize the physical implementation of quantum gates, using techniques like randomized compiling (RC). In this work, we use gate set tomography to perform precision characterization of a set of two-qubit logic gates to study RC on a superconducting quantum processor. We find that, under RC, gate errors are accurately described by a stochastic Pauli noise model without coherent errors, and that spatially-correlated coherent errors and non-Markovian errors are strongly suppressed. We further show that the average and worst-case error rates are equal for randomly compiled gates, and measure a maximum worst-case error of 0.0197(3) for our gate set. Our results show that randomized benchmarks are a viable route to both verifying that a quantum processor's error rates are below a fault-tolerance threshold, and to bounding the failure rates of near-term algorithms, if --- and only if --- gates are implemented via randomization methods which tailor noise.
\end{abstract}

\keywords{QCVV, Fault Tolerance, Gate Set Tomography, Randomized Compiling}

\maketitle

\section*{Introduction}\label{sec:introducton}
\noindent
Quantum bits (qubits) in the noisy intermediate-scale quantum (NISQ) \cite{preskill2018quantum} era are short-lived and susceptible to a variety of errors and noise due to imperfect control signals and imperfect isolation from the surrounding environment. Therefore, utilizing quantum computers to solve classically-intractable problems (e.g., integer factoring \cite{shor1999polynomial}) will likely require quantum error correction (QEC) \cite{shor1995scheme, gottesman1996class, steane1996multiple, steane1998introduction, calderbank1998quantum}. QEC can protect logical qubits from errors, but it is only guaranteed to work if the error rate of each physical qubit is below some fault tolerance (FT) threshold \cite{shor1996fault, knill1996concatenated, knill1998resilient, preskill1998fault, kitaev2003fault, aharonov2008fault}. Analytic lower bounds on FT thresholds for various QEC codes have been derived, ranging from $\sim \! 10^{-6}$ for generic local noise \cite{aharonov2008fault} to $\sim \! 10^{-5}$--$10^{-3}$ for stochastic and depolarizing noise \cite{aliferis2005quantum, aliferis2007subsystem, aliferis2007accuracy, aliferis2009fibonacci, chamberland2016thresholds}. More optimistic estimates obtained via numerical simulation are orders of magnitude larger than the lower bounds, ranging from $\sim \! 10^{-3}$--$10^{-1}$ \cite{knill2005quantum, aliferis2009fault, duclos2010fast, wang2011surface, bombin2012strong, wootton2012high, stephens2014fault, auger2017fault, tuckett2020fault}, but often assume stochastic (e.g., Pauli, dephasing, or depolarizing) noise models. While recent claims of quantum gates approaching or surpassing FT thresholds boast impressive gate fidelities \cite{barends2014superconducting, rong2015experimental, xue2022quantum}, there is a discrepancy between these claims and the formulation of FT thresholds, which are specified in terms of worst-case error rates.

Various error metrics and measures exist for quantifying the ``error rate'' of a quantum gate. Randomized benchmarks \cite{emerson2005scalable, knill2008randomized, dankert2009exact, magesan2011scalable, magesan2012efficient} typically define error rates in terms of the average gate fidelity, or, equivalently, the process fidelity
\begin{equation}
    F = \bra{\psi} (\mathbb{I} \otimes \mathcal{E})(\rho) \ket{\psi},
\end{equation}
where $\rho =\ketbra{\psi}$ is a maximally-entangled state, $\mathcal{E}$ is the error channel associated with some quantum gate, and $\mathbb{I}$ the identity operation. A gate's process \textit{infidelity} $e_F(\mathcal{E}) = 1 - F$ (or \emph{average error rate}) quantifies the probability that the gate induces an error on a random input state, or, equivalently, the average failure rate of random circuits that contain one instance of this gate but that are otherwise perfect. However, FT thresholds are typically defined via each gate's worst-case error rate (also called the diamond norm) \cite{kitaev1997quantum},
\begin{equation}\label{eq:diamond_norm}
    \epsilon_\diamond(\mathcal{E}) = \frac{1}{2} \big|\big| \mathcal{E} - \mathbb{I} \big|\big|_\diamond = \frac{1}{2} \sup_{\rho} \big|\big| \left[ \mathbb{I} \otimes (\mathcal{E} - \mathbb{I}) \right](\rho) \big|\big|_1,
\end{equation}
where the supremum is taken over all pure states and $\big|\big| X \big|\big|_1 = \textrm{Tr} \sqrt{X^\dagger X}$ is the trace norm. Operationally, $\epsilon_\diamond(\mathcal{E})$ represents the worst-case performance of a quantum gate in any circuit, whereas $e_F(\mathcal{E})$ represents the average-case performance for a single instance of the gate. While the diamond norm is a pessimistic estimate of the error rate of a quantum gate, it provides much more rigorous performance guarantees in the context of QEC. This is because the diamond norm upper bounds the accumulation of error in any quantum circuit, since the distance between the ideal and actual output probability distributions (measured via total variation distance) for any circuit is bounded above by the sum of the worst-case error rates of all its gates \cite{kitaev1997quantum}.

While $e_F(\mathcal{E})$ can be measured directly via randomized benchmarks, there exists no known scalable method for measuring $\epsilon_\diamond(\mathcal{E})$. Tomographic methods, such as gate set tomography (GST) \cite{merkel2013self, blume2013robust, greenbaum2015introduction, blume2017demonstration, nielsen2021gate}, can be used to estimate $\epsilon_\diamond(\mathcal{E})$ \cite{blume2017demonstration}, but they are exponentially expensive in the number of qubits. If only $e_F(\mathcal{E})$ is known, $\epsilon_\diamond(\mathcal{E})$ can be bounded using \cite{wallman2014randomized, sanders2015bounding, wallman2015bounding, kueng2016comparing}
\begin{equation}\label{eq:bounds}
    e_F(\mathcal{E}) \leq \epsilon_\diamond(\mathcal{E}) \leq \sqrt{e_F(\mathcal{E})}d,
\end{equation}
where $d = 2^n$ (for $n$ qubits). The lower bound of $\epsilon_\diamond(\mathcal{E})$ is saturated when $\mathcal{E}$ is a stochastic Pauli channel, and the upper bound, which is quadratically larger in $e_F$ and scales with the dimension $d$, is saturated by a unitary channel. While modern experimental platforms routinely report single- and two-qubit infidelities on the order of $e_{F,1Q} \lesssim 10^{-4}$ and $e_{F,2Q} \lesssim 10^{-2}$ \cite{sheldon2016characterizing, arute2019quantum, wang2020high, pino2021demonstration, mitchell2021hardware, xue2022quantum, mkadzik2021precision}, respectively, even if coherent errors account for a tiny fraction of the infidelity, they can dominate the diamond norm, in which case worst-case error rates can be as large as $\epsilon_{\diamond,1Q} \sim \sqrt{e_{F,1Q}} \lesssim 10^{-2}$ and $\epsilon_{\diamond,2Q} \sim \sqrt{e_{F,2Q}} \lesssim 10^{-1}$ \cite{sanders2015bounding, wallman2015bounding, wallman2016noise}. Therefore, $e_F(\mathcal{E})$ and $\epsilon_\diamond(\mathcal{E})$ can differ by orders of magnitude in the presence of coherent errors. This means that randomized benchmarks are generally inadequate for testing whether gate error rates are below FT thresholds \cite{sanders2015bounding, kueng2016comparing}. Refs.~\cite{barends2014superconducting, rong2015experimental, xue2022quantum} report single- and two-qubit error rates below the FT threshold for the surface code, but base their claims on average error rates from randomized benchmarking (RB) or GST, not diamond norms. Refs.~\cite{ryan2021realization, chen2023transmon} make similar claims about state-preparation and measurement (SPAM) errors, but only report average preparation and assignment fidelities, not their worst-case infidelities. Notably, Ref.~\cite{xue2022quantum} includes estimates of the diamond norm that are an order of magnitude larger than their reported process infidelities.

While it is not generally possible to directly compare $e_F(\mathcal{E})$ to a FT threshold for QEC, if it can be guaranteed that $e_F(\mathcal{E}) \approx \epsilon_\diamond(\mathcal{E})$ then randomized benchmarks can be used to efficiently verify that gate error rates are below a FT threshold. One method for ensuring that an error budget is dominated by stochastic noise is randomized compiling (RC) \cite{wallman2016noise, hashim2021randomized}, which converts all gate errors into stochastic Pauli channels via Pauli twirling. This ensures that the direct measurement of $e_F(\mathcal{E})$ (e.g., via cycle benchmarking \cite{erhard2019characterizing}) accurately captures the worst-case error rate, which enables the comparison to FT thresholds as well as bounding the overall failure rate of any quantum circuit or application.

In this work, we use GST to study RC performed on two qubits (labeled Q5 and Q6; see Methods) on a superconducting transmon processor ($\texttt{AQT@LBNL Trailblazer8-v5.c2}$). GST enables measurements of both the process infidelity and diamond norm for all gates in our gate set, allowing us to study the impact of RC on gate errors. We find that RC eliminates signatures of coherent errors, enabling one to accurately describe the gates' errors by stochastic Pauli noise. We further show that RC suppresses spatially-correlated coherent errors and non-Markovian errors. Finally, we show that the diamond norm converges to the process infidelity under RC, saturating the lower bound of Eq.~\ref{eq:bounds}, providing strong experimental evidence that our quantum logic operations are approaching or below a threshold for fault tolerance. By combining RC with GST, our results provide a novel framework for verifying that FT-required assumptions are satisfied, demonstrating that FT thresholds can be accurately measured using randomized benchmarks as long as quantum circuits are implemented using RC or related randomization methods \cite{hastings2016turning, campbell2017shorter, cai2020mitigating, ware2021experimental, hashim2021optimized}.
\section*{Results}\label{sec:results}

\paragraph*{\textbf{\textup{Gate Set Tomography.}}}\label{sec:gst}

\begin{figure*}[t]
    \centering
    \includegraphics[width=2\columnwidth]{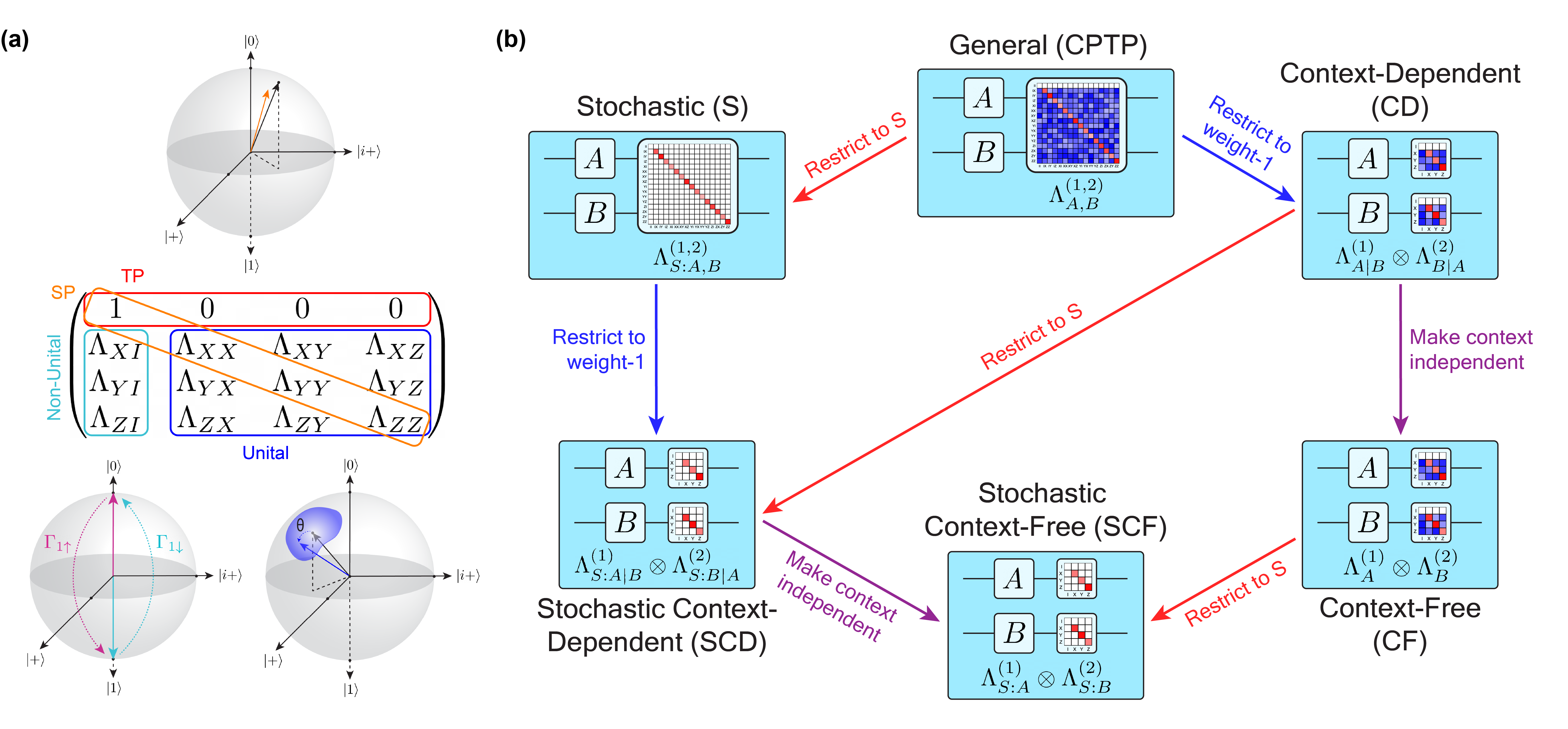}
    \caption{\textbf{Modeling gate errors in the Pauli transfer matrix (PTM) superoperator representation.} \textbf{(a)} The PTM is divided into four blocks: the top row (red) represents trace-preservation (TP). The lower right-hand block (blue) captures unital processes, such as unitary errors. The leftmost column (cyan) captures non-unital processes, such as $T_1$ decay. And the diagonal of the PTM (orange) represents state preservation (SP); $\Lambda_{PP} = 1$ ($< 1$) if the Pauli channel $P$ is (not) preserved under an error process such as stochastic Pauli noise.
    \textbf{(b)} Hierarchy of nested GST models. For single-qubit gates $A$ and $B$ acting on qubits 1 and 2, respectively, the general CPTP model contains all weight-1 and weight-2 error generators. The S model restricts the errors to be stochastic (i.e., the error generator is diagonal in the PTM) [red arrows]. The CD (SCD) model restricts the error generators to be weight-1 (stochastic) errors [blue arrows], but allows contextual dependence. The CF (SCF) model restricts the error generators to be context-independent weight-1 (stochastic) errors [purple arrows].}
    \label{fig:figure1}
\end{figure*}

\noindent
Gate set tomography is a robust method for tomographically reconstructing the errors and noise impacting all gate operations within a gate set \cite{nielsen2021gate}. Like traditional quantum process tomography (QPT) \cite{chuang1997prescription}, GST fully characterizes the process matrix of a quantum gate; however, unlike QPT, it does so in a self-consistent manner that simultaneously characterizes the errors in all the gates and in the state-preparation and measurement (SPAM). Using the open-source Python library \texttt{pyGSTi} \cite{nielsen2019python, nielsen2020probing}, one can obtain a best-fit model for the gate set, consisting of a process matrix for each gate, using maximum-likelihood estimation \cite{blume2017demonstration, nielsen2021gate}. In this work, our gate set consists of all possible combinations of $I$, $X_{\pi/2}$, and $Y_{\pi/2}$ single-qubit gates applied simultaneously to both qubits, as well as the controlled-Z ($CZ$) gate: $\mathbb{G} = \{G_1 \otimes G_2: G_1, G_2 \in \{I, X_{\pi/2}, Y_{\pi/2}\}\} \cup \{CZ\}$. We utilized GST up to depth $L = 128$ layers to benchmark the performance of all gates in the gate set.

In this work, we represent gate errors using the Pauli transfer matrix (PTM) representation of superoperators (see Fig.~\ref{fig:figure1}a and Methods), denoted $\Lambda$. We fit our data to the following parameterized error models:
\begin{enumerate}
    \item General completely-positive and trace-preserving (CPTP) model: each gate's errors are modeled by a general CPTP two-qubit PTM $\Lambda^{(1,2)}_{A,B}$, where $A$ ($B$) denotes the gate acting on qubit 1 (2).
    \item General Pauli stochastic (S) model: the errors in the general model are restricted to be diagonal in the Pauli basis $\left( \Lambda^{(1,2)}_{S:A,B} \right)$.
    \item Context-dependent (CD) model: the errors on each single-qubit gate is described by a single-qubit PTM that can depend on the gate acting on the other qubit $\left( \Lambda^{(1)}_{A|B} \otimes \Lambda^{(2)}_{B|A} \right)$.
    \item Stochastic context-dependent (SCD) model: the errors in the CD model are restricted to be stochastic Pauli errors $\left( \Lambda^{(1)}_{S:A|B} \otimes \Lambda^{(2)}_{S:B|A} \right)$.
    \item Context-free (CF) model:  the errors on each single-qubit gate is described by a single-qubit PTM that is unconditional on the gate acting on the other qubit $\left( \Lambda^{(1)}_{A} \otimes \Lambda^{(2)}_{B} \right)$.
    \item Stochastic context-free (SCF) model: the errors in the CF model are restricted to be stochastic Pauli errors $\left( \Lambda^{(1)}_{S:A} \otimes \Lambda^{(2)}_{S:B} \right)$.
\end{enumerate}
The hierarchy of all of the nested GST models can be seen in Fig.~\ref{fig:figure1}b.

In general, models with greater complexity are able to capture more complex dynamics; however, they also require fitting a larger number of free parameters. The general CPTP model makes no assumption of locality and can completely capture all two-qubit interactions, including quantum crosstalk errors. The CD model assumes that errors on single-qubit gates must be local (weight-1), since the PTMs decompose as a tensor product of operations, but allows the errors to be classically correlated (i.e., the error impacting qubit 1 can be correlated with the gate being applied to qubit 2, but it cannot depend on the state of qubit 2). This can model errors due to classical crosstalk, but not entangling quantum crosstalk. The CF model assumes that the errors on single-qubit gates are local and independent of the gate being applied to the other qubit. The corresponding S-type models make the same assumptions, but restrict the errors to be stochastic Pauli noise. All models make an assumption of Markovianity.
\newline

\paragraph*{\textbf{\textup{Randomized Compiling.}}} \label{sec:rc}

\begin{figure*}[ht]
    \centering
    \includegraphics[width=2\columnwidth]{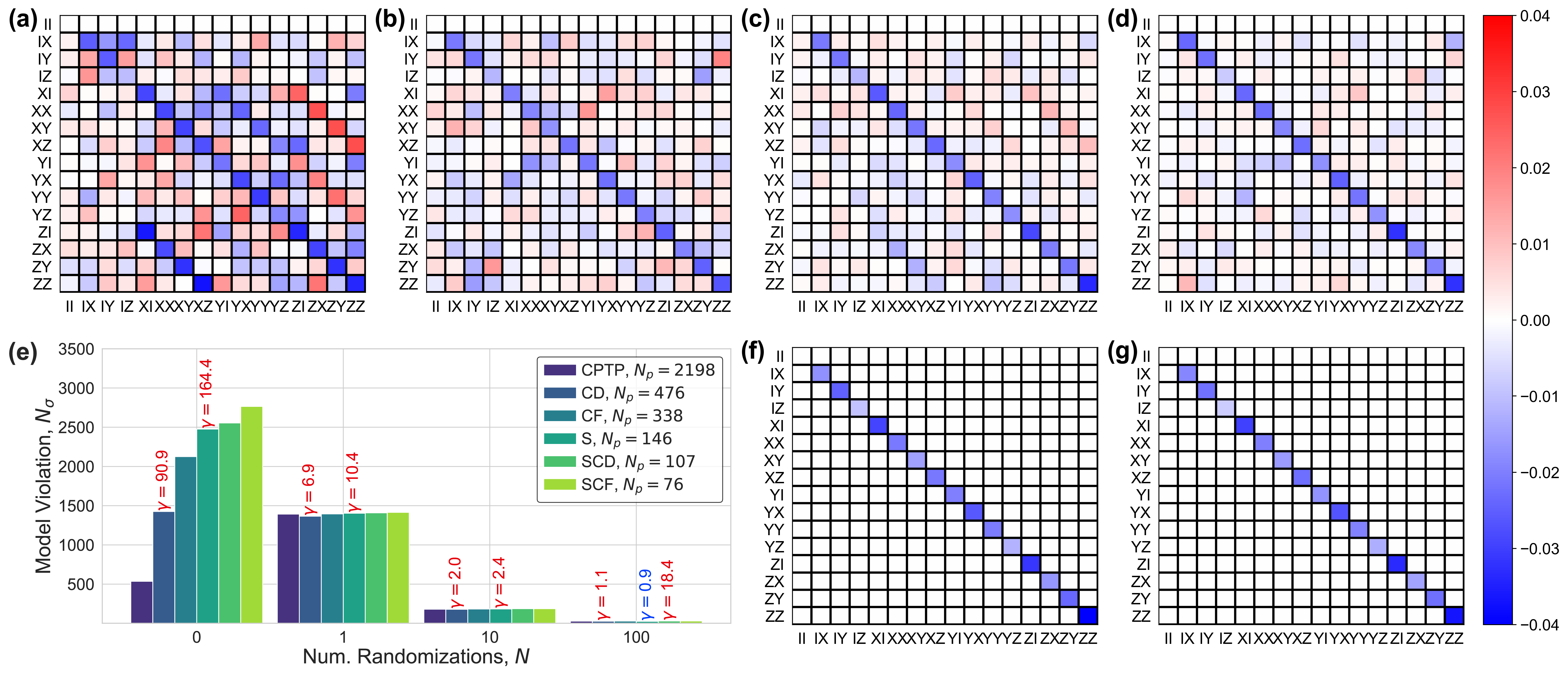}
    \caption{\textbf{Improving model accuracy via noise tailoring.} 
    The error generator $\mathcal{L}$ in the PTM representation for the CPTP model of the $CZ$ gate is plotted for \textbf{(a)} $N=0$, \textbf{(b)} $N=1$, \textbf{(c)} $N=10$, and \textbf{(d)} $N=100$ randomizations under RC. If an error channel (i.e., PTM cell) in $\mathcal{L}$ is zero, then this component of the estimated gate matches the corresponding component of the ideal target gate.
    \textbf{(e)} The model violation $N_\sigma$ for each GST model, plotted as a function of $N$. The evidence ratio $\gamma$ is labeled above each nested model, and the number of free parameters $N_p$ of each model is listed in the legend. For each $N$, we choose the model with the least number of parameters that satisfies $\gamma \le 1$, finding that CPTP is the best fit for $N=0$, $N=1$, and $N=10$, and S for $N=100$; blue (red) text indicates that the model is accepted (rejected). The $CZ$ gate error generator $\mathcal{L}$ is plotted for the S model for \textbf{(f)} $N=10$ and \textbf{(g)} $N=100$, showing that the stochastic models capture the dominant error for both. Even though it is ultimately rejected by the evidence ratio for $N=10$, the S model could be reasonably selected for its simplicity. As the S model is selected for $N=100$, all off-diagonal elements in \textbf{d} are statistically consistent with zero.}
    \label{fig:figure2}
\end{figure*}

\noindent
Randomized compiling is an efficient method for converting arbitrary Markovian errors into stochastic Pauli noise for each cycle (or layer) in a quantum circuit via Pauli twirling. This can benefit circuit performance in two different regimes (see Methods):
\begin{enumerate}
    \item Single-randomization limit: a single randomization under RC interrupts the coherent accumulation of unitary errors between cycles of gates, similar to dynamical decoupling sequences \cite{viola1999dynamical}.
    \item Many-randomization limit: averaging over many ($N$) randomizations under RC tailors errors into stochastic Pauli channels, completely eliminating off-diagonal terms in the error process in the limit that $N \longrightarrow \infty$.
\end{enumerate}
In this work, we apply RC to GST circuits using $N=1$, $N=10$, and $N=100$ randomizations to study the transition of RC from the single- to many-randomization limit, and compare the results to GST performed without randomization (denoted by $N=0$).

To study the types of errors present in a gate's process matrix $G$ estimated using GST, we write a noisy quantum gate as
\begin{equation}
    G = \Lambda G_0 = e^\mathcal{L} G_0,
\end{equation}
where $G_0$ is the ideal quantum gate, $\Lambda$ the gate error channel, and $\mathcal{L}$ the gate error generator \cite{blume2022taxonomy}. $\mathcal{L}$ is analogous to the Linbladian superoperator that generates all gate errors (coherent, stochastic, and non-unital).

Fig.~\ref{fig:figure2} shows the error generator $\mathcal{L}$ of the $CZ$ gate for $N = 0, 1, 10, 100$ randomizations. We find that RC transforms the error generator from dense to sparse as we increase $N$, twirling $\mathcal{L}$ into a stochastic Pauli channel (these channels correspond to diagonal PTMs). Going from $N=0$ (Fig.~\ref{fig:figure2}a) to $N=1$ (Fig.~\ref{fig:figure2}b) randomizations significantly reduces the magnitude of $\mathcal{L}$'s off-diagonal elements, which signifies the presence of coherent errors, but $\mathcal{L}$ for $N=1$ still has significant magnitude in its off-diagonal elements. As $N$ increases from 1 to 10, and from 10 to 100, the magnitude of the off-diagonal terms is greatly reduced. This is referred to as the ``noise tailoring'' property of RC.
\newline
\paragraph*{\textbf{\textup{Model Selection.}}}\label{sec:model_selection}

\noindent
GST enables the comparison of nested error models in a self-consistent manner \cite{nielsen2021efficient, rudinger2021experimental}. To compare two nested models, we utilize the \textit{evidence ratio} $\gamma$ \cite{nielsen2021efficient},
\begin{equation}
    \gamma = \frac{\lambda_S - \lambda_L}{N_{p,L} - N_{p,S}},
\end{equation}
where $\lambda$ is the model's log-likelihood ratio (see Methods), and L (S) denotes the larger (smaller) model defined in terms of the number of free parameters $N_p$ describing the model. If $\gamma \le 1$, then we automatically choose the smaller model, as it describes the data at least as well as the larger model without extra (potentially unused) parameters. If $\gamma > 1$, there is evidence for rejecting the smaller model, but even if $1 \lesssim \gamma \lesssim 10$ the smaller model may still be preferable due to its simplicity \cite{mkadzik2021precision}. In this work, we compute the evidence ratio for each pair of nested models down the two separate branches of the hierarchy in Fig.~\ref{fig:figure1}b: if $\gamma \le 1$ we select the smaller model and continue down the hierarchy until $\gamma > 1$, at which point we select the larger of the two models and stop. If two independent models satisfy this criteria, we select the model with the fewest free parameters, as this model represents the best fit estimate of our data without over-fitting.

Additionally, we compute the model violation $N_\sigma$ of each model, which captures the number of standard deviations that $\lambda$ is from the expected mean (see Methods). If $N_\sigma \le 1$, then the GST model faithfully captures all of the errors in the device. However, on actual NISQ hardware, typically $N_\sigma > 1$ (and sometimes even $N_\sigma \gg 1$) is observed \cite{blume2017demonstration, ware2021experimental, rudinger2021experimental}, indicating the presence of significant model violation. Because GST can model all Markovian gate errors, large $N_\sigma$ for the CPTP model indicates that there is strong evidence in the data for non-Markovian errors --- which can be true even if these non-Markovian errors are small in magnitude (see the Non-Markovian Errors section for further discussion).

To choose a best-fit model for each dataset, we compute $\gamma$ and $N_\sigma$ for each model, shown in Fig.~\ref{fig:figure2}e. We find that the general CPTP model fits best for $N=0$, $N=1$, and $N=10$, and that the S model fits as well as the general CPTP model for $N=100$. For $N=10$, none of the nested models satisfy the evidence ratio criteria, but the S or CD model could also be reasonably selected, as $\gamma=2.0$ and $2.4$, respectively (there is therefore only weak evidence for coherent errors). For $N=100$, only the S model satisfies the evidence ratio criteria, and we therefore prefer this model over the general CPTP model. For the data presented in the rest of this work, we fix the model for each dataset using the best-fit models outlined above.

The large model violation for $N=0$ ($N_\sigma \approx 535$) is strong evidence that there are non-Markovian errors, which cannot be modeled by a PTM. $N=10$ ($N_\sigma \approx 180$) and $N=100$ ($N_\sigma \approx 25$) have significantly less model violation, indicating less statistical evidence of non-Markovian errors. For a single randomization ($N=1$) there is even larger model violation ($N_\sigma \approx 1394$) than with no randomization ($N=0$).
This effect arises because when $N=1$, each circuit is replaced with a single randomization of that circuit, and no averaging occurs. Therefore, the implementation of a logical gate (in $\mathbb{G}$) will not correspond to the average action of multiple PTMs (which decoheres errors) corresponding to the different Pauli-equivalent implementations of the gate, but instead the action of a particular PTM corresponding to one randomly selected Pauli-equivalent implementation of the gate. Importantly, the PTM describing the action of a logic gate will change depending on the randomization selected (and therefore it varies between circuits and uses of the logical gate within a circuit). Therefore, even when each physical gate is describable by a PTM (i.e., all errors are Markovian), when $N=1$ each logical gate is not generally describable by a single PTM. This violates the assumptions of GST --- which finds the best-fit PTM for each gate --- and so a GST model fit to the $N=1$ data can exhibit significant model violation.
(See Methods for further discussion and examples.) \\

\paragraph*{\textbf{\textup{Impact of Randomized Compiling on Error Budgets.}}}\label{sec:error_budget}

\begin{figure*}[th]
    \centering
    \includegraphics[width=2.0\columnwidth]{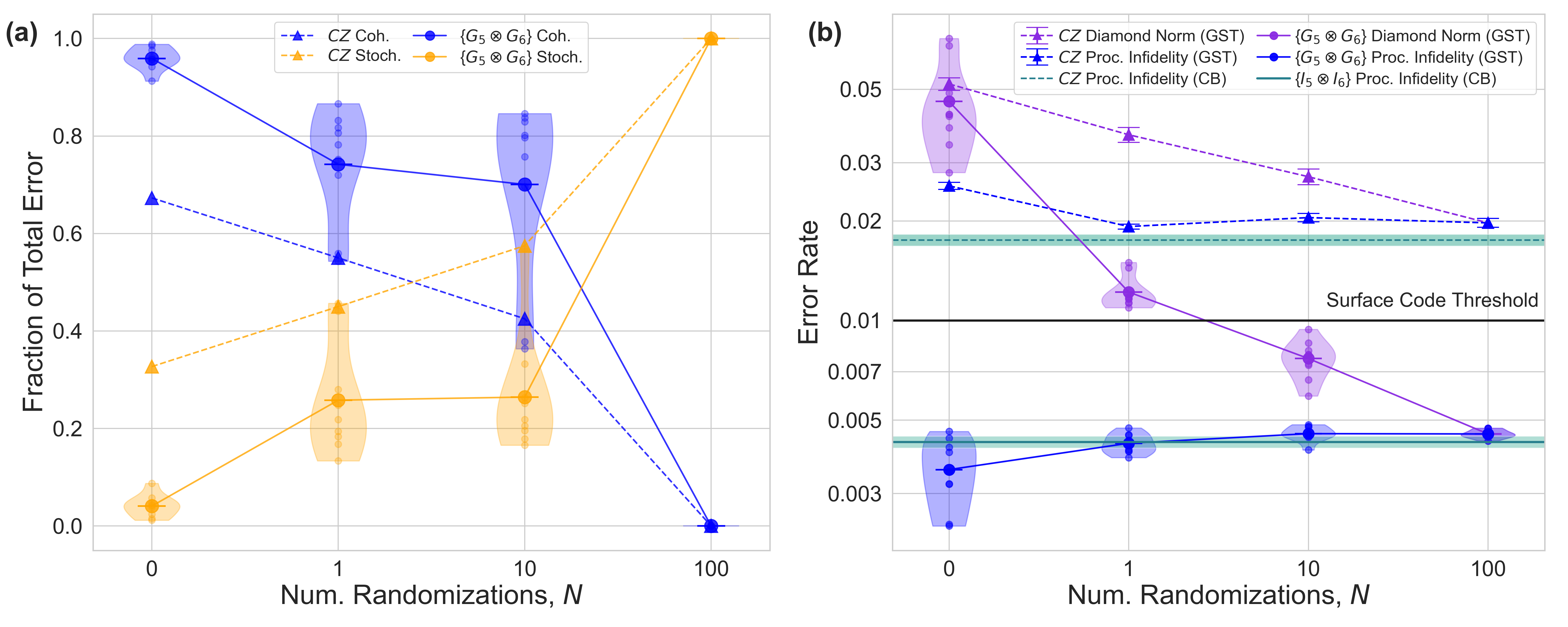}
    \caption{\textbf{Tailoring noise for improved estimates of the diamond norm.} 
    \textbf{(a)} Fraction of the total error due to coherent errors (blue) and stochastic noise (orange) as a function of the number of randomizations $N$ under RC. As $N$ increases, stochastic noise makes up a larger fraction of the total error, which is dominated by coherent errors for small $N$. For \textbf{a} and \textbf{b}, triangular markers denote the $CZ$ gate, and circular markers denote the single-qubit gates $\{G_5 \otimes G_6\}$ (small transparent markers depict the individual gates, large  markers depict the averages, and violin plots outline the distribution). 
    \textbf{(b)} Saturation of the lower bound of the diamond norm under RC. We plot the GST process infidelity $e_F$ (blue) and diamond norm $\epsilon_\diamond$ (purple) as a function of the number of randomizations $N$. We observe that $\epsilon_\diamond > e_F$ for all gates for $N=0,1,10$, but that the two are equal for $N=100$. For visual clarity, we omit the idle cycle for $N=0$, as $e_{F,II} = 0.0001$ and $\epsilon_{\diamond,II} = 0.001$ fall well below their respective averages. We compare the GST estimates with the process infidelity measured via cycle benchmarking (CB) for the $CZ$ gate (dashed green line) and idle cycle $\{I_5 \otimes I_6\}$ (solid green line). The GST error bars and CB transparent bands indicate the 95\% confidence intervals. Additionally, we compare the benchmarked error rates with the 1\% FT threshold for the surface code (black line) and find that the single qubit gates are well below the threshold value, whereas the $CZ$ gate is approaching, but does not surpass the threshold.}
    \label{fig:figure3}
\end{figure*}

\noindent
To explore how effective RC is at converting all errors into stochastic Pauli noise, we calculate the total amount of stochastic and coherent errors for each gate. To do so, we divide each gate's error generator $\mathcal{L}$ into stochastic and Hamiltonian components, and compute the \emph{total error},
\begin{equation}
    \epsilon_{\rm{tot}} = \epsilon_{\rm{agg}}+ \theta_{\rm{agg}} ,
\end{equation}
where $\epsilon_{\rm{agg}} = \sum_i \epsilon_i$ is the sum of the rates of all stochastic error generators, and $\theta_{\rm{agg}} = \sqrt{\sum_i \theta^2_i}$ is the quadrature sum of all Hamiltonian error generators \cite{mkadzik2021precision}. The total error is closely related to the diamond norm. For single qubit error generators $\mathcal{L}$, the total error upper-bounds the diamond norm error, $\epsilon_\diamond\left(e^\mathcal{L}\right) \le \epsilon_{\rm{tot}}\left(e^\mathcal{L}\right)$. For two or more qubits, the bound is much weaker \cite{mkadzik2021precision}.

In Fig.~\ref{fig:figure3} we plot the fraction of the total error due to stochastic and coherent (Hamiltonian) errors. We find that the error budget of the simultaneous single-qubit gates is dominated by coherent errors for $N=0$, and that coherent errors account for approximately two-thirds of the total error for the $CZ$. For $N=1$, coherent errors still dominate the single-qubit gates, but the contribution from coherent errors and stochastic noise are both approaching 50\% for the $CZ$ gate. By $N=10$, stochastic noise makes up the largest contribution to the total error for the $CZ$ gate, but we observe only a modest change for the single-qubit gates. However, by $N=100$ the error budget for all gates is entirely due to stochastic noise. We note that for the $N=100$ data, our model selection chose the S model, which enforces the constraint that the error budget is entirely due to stochastic Pauli noise (see the diagram in Fig.~\ref{fig:figure1}b). This is because there is no statistically significant evidence in the data for any coherent errors (if more data were taken, evidence for some residual coherent errors might be found). These data demonstrate the effectiveness of RC in eliminating the impact of coherent errors. However, this does not mean that coherent errors are physically not present; rather, each individual randomization under RC is impacted by coherent errors in a different manner, in such a way that the aggregate effect is to randomize the impact of coherent errors.

Alternative methods for quantifying the magnitude of coherent errors include purity RB (PRB) \cite{wallman2015estimating} and cross-entropy benchmarking (XEB) \cite{boixo2018characterizing}. However, methods for measuring coherent errors that use random circuits have two disadvantages. First, these methods measure averages over gates, so they cannot separate out each gate's error rate into stochastic and coherent contributions (as done herein). Second, although the unitarity (estimated by PRB) and infidelity (estimated by RB) can be used to upper bound the diamond norm error \cite{wallman2015estimating}, 
this method is inefficient (see Ref.~\cite{blume2017demonstration} for details).
\newline
\paragraph*{\textbf{\textup{Process Infidelity vs.~Diamond Norm.}}}\label{sec:infidelity_diamon_norm}

\noindent
Error rates measured via randomized benchmarks cannot in general be directly compared to FT thresholds. However, although the diamond norm (Eq.~\ref{eq:diamond_norm}) enables rigorous comparison to FT thresholds, it is difficult to measure in a scalable manner. If the error model of a quantum processor is dominated by stochastic noise, the process infidelity and diamond norm will be equal, in which case randomized benchmarks can be used to efficiently demonstrate that gate errors are below a FT threshold.

In Fig.~\ref{fig:figure3}, we plot the process infidelity and the diamond norm as a function of $N$ for all gates in our gate set. We find that $\epsilon_\diamond(\mathcal{E})$ converges to $e_F(\mathcal{E})$ as $N$ increases. This is strong experimental evidence that $\epsilon_\diamond(\mathcal{E}) = e_F(\mathcal{E})$ in the many-randomization limit for $N=100$ (saturating the lower bound of Eq.~\ref{eq:bounds}). We also compare these results with process infidelities measured independently via CB and find good agreement between the two. These results demonstrate that randomized benchmarks are sufficient for benchmarking FT thresholds if --- and only if --- a quantum application is impacted only by stochastic noise, which can be guaranteed if implemented using methods which tailor noise. We note that similar results were previously reported using Pauli frame randomization \cite{knill2004fault, kern2005quantum} for single-qubit gates \cite{ware2021experimental}.

The largest diamond norm error over a gate set is arguably the most relevant quantity to compare to a FT threshold. The surface code \cite{fowler2012surface} is a popular QEC code due to its high FT threshold, which is estimated to be between $\sim 0.75\% - 3\%$ \cite{wang2003confinement, raussendorf2006fault, raussendorf2007topological, raussendorf2007fault, wang2011surface, fowler2012towards}, with 1\% being the threshold that is typically quoted in the literature \cite{barends2014superconducting, xue2022quantum}. In Fig.~\ref{fig:figure3}, we find that the diamond norm of simultaneous single-qubit gates is below the surface code threshold for $N=10$ and $N=100$, and that the error rate of our $CZ$ gate is approaching --- but does not surpass --- the surface code threshold.
\newline
\paragraph*{\textbf{\textup{Correlated Errors Under Pauli Twirling.}}}\label{sec:corr_errors}

\begin{figure*}[ht]
    \centering
    \includegraphics[width=2\columnwidth]{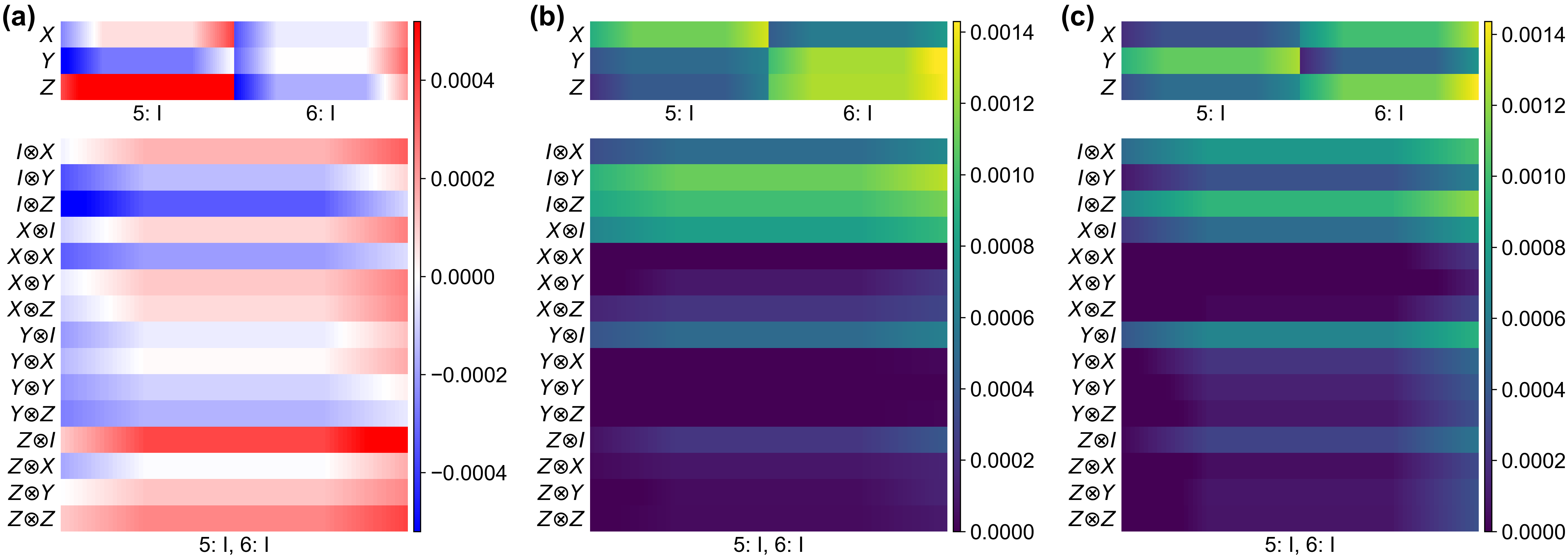}
    \caption{\textbf{Suppressing correlated errors.} Heat maps of the weight-1 and weight-2 errors acting on Q5 and Q6 during the idle cycle $\{I_5 \otimes I_6\}$ reconstructed using \textbf{(a)} the Hamiltonian projection of the GST error generator $\mathcal{L}$ for $N=0$, \textbf{(b)} error rates calculated from the stochastic projection of $\mathcal{L}$ for $N=100$, and \textbf{(c)} cycle error reconstructing via CB measurements.
    The x-axis labels the target gate; the y-axis labels the Hamiltonian error for \textbf{a}, and the Pauli Kraus error for \textbf{b} and \textbf{c}; the cell color denotes the over-rotation angle for \textbf{a}, and error rate for \textbf{b} and \textbf{c}; the cell gradient defines the 95\% confidence interval; the first (second) row of subplots shows marginalized weight-1 (correlated weight-1 and weight-2) errors.
    While weight-2 errors are dominant for $N=0$, \textbf{b} and \textbf{c} show that Pauli twirling suppresses weight-2 errors to negligible levels.}
    \label{fig:figure4}
\end{figure*}

\noindent
A major requirement for reliable fault-tolerant QEC is the absence of correlated errors, which can occur temporally \cite{lu2007quantum} or spatially \cite{wilen2021correlated}. Many-qubit correlated errors cannot be corrected by QEC (each QEC scheme has a maximal weight of error that it can correct), causing logical failures. Therefore, the rate of correlated errors must be low to achieve reliable fault-tolerant quantum computation. To characterize the extent to which spatially correlated errors are present in our system with and without Pauli twirling, we extract the weight-1 and weight-2 errors of each gate in our gate set. 
Fig.~\ref{fig:figure4} shows the weight-1 and weight-2 coherent (stochastic) errors for the idle cycle $\{I_5 \otimes I_6\}$ for the CPTP (S) model for $N=0$ ($N=100$) randomizations. We focus on coherent (stochastic) errors for $N=0$ ($N=100$) as they dominate the total the error budget; see Fig.~\ref{fig:figure3}a. We observe significant weight-2 coherent errors for $N=0$ (Fig.~\ref{fig:figure4}a), which corresponds to unintended entanglement (i.e., quantum crosstalk), such as static $ZZ$ coupling \cite{mundada2019suppression, zhao2020high, ni2021scalable}. In contrast, we observe that weight-1 Pauli errors dominate for $N=100$, and that weight-2 errors are largely suppressed in comparison.

Additionally, we compare the Pauli error rates for $N=100$ (Fig.~\ref{fig:figure4}b) to independently estimated Pauli error rates (Fig.~\ref{fig:figure4}c) measured using cycle benchmarking (CB) \cite{erhard2019characterizing} and reconstructed using cycle error reconstruction (CER) \cite{flammia2020efficient, hashim2021randomized, beale_stefanie_j_2020_3945250}. We find good agreement between the reconstructed Pauli noise in the two error maps, showing similar magnitudes of correlated errors, and demonstrating that the dominant errors in our system are weight-1. These results show that that Pauli twirling can suppress spatially-correlated coherent errors due to entangling crosstalk and static $ZZ$ coupling between superconducting qubits. (See Methods for a similar CER analysis of correlated errors in cycles acting on four qubits.)

In general, one can expect RC to provide a quadratic suppression of correlated coherent errors --- RC converts a coherent error that contributes $\mathcal{O}(\theta)$ to the diamond norm (and the total error) into a stochastic error that contributes $\mathcal{O}(\theta^2)$ to the diamond norm. Therefore, while correlated errors might not be entirely suppressed by Pauli twirling, this reduction in the magnitude of correlated errors may reduce the cost of QEC \cite{fowler2014quantifying}. While the current work focuses on spatial correlations, temporal correlations can also be detrimental to worst-case error rates \cite{ball2016effect}. However, we expect that a single randomization under RC can similarly suppress temporally correlated errors, although we leave an exploration of this topic to future work.
\newline
\paragraph*{\textbf{\textup{Non-Markovian Errors.}}}\label{sec:nm_errors}

\begin{figure*}[ht]
    \centering
    \includegraphics[width=2\columnwidth]{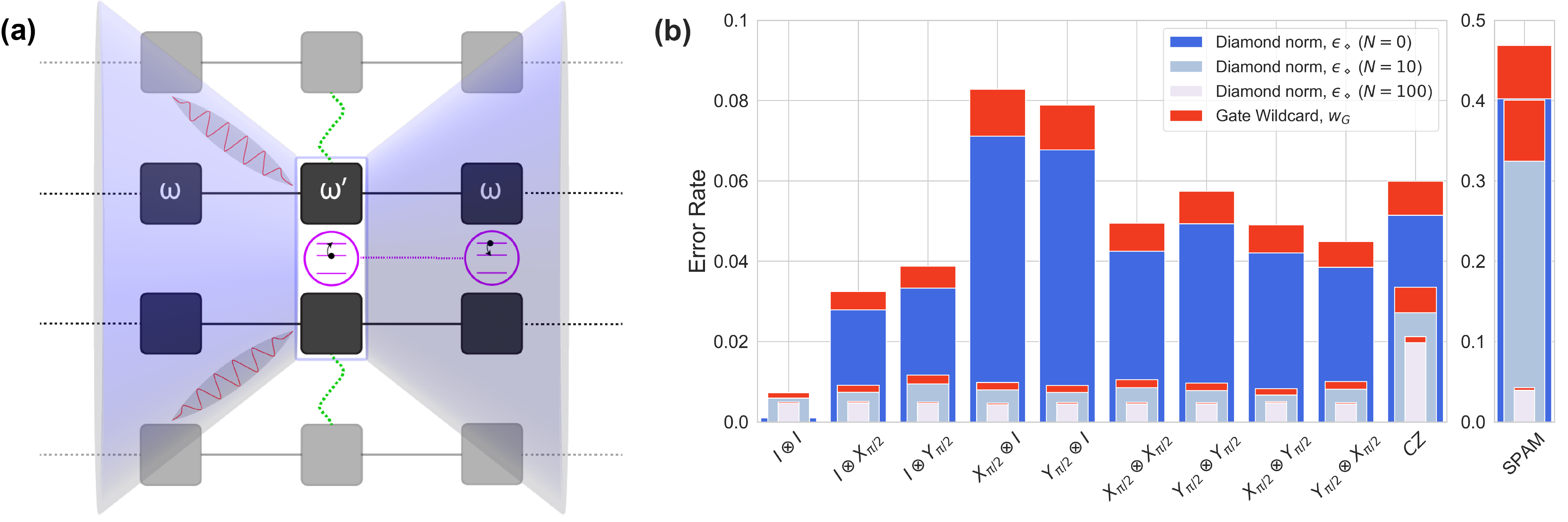}
    \caption{\textbf{Suppressing non-Markovian errors.}
    \textbf{(a)} Non-Markovian errors in gate-based quantum computing. Markovian errors for a given system of qubits (black) are defined to occur within the timescale of a given cycle of gates (blue rectangle). Sources of non-Markovianity include leakage (seapage) out of (into) the computational basis states (purple), unintended entanglement (green) with external qubits (grey) or two-level fluctuators in the environment, drift in the system properties (e.g., fluctuations in the qubit frequency $\omega$, with $\omega' = \omega + \delta\omega$), and classical EM signals from outside of the defined system (red) that reach the system qubits within their pseudo light cone (blue).
    \textbf{(b)} Unmodeled error versus the diamond norm. The per-gate wildcard $w_G$ is plotted on top of the diamond norm for each gate in $\mathbb{G}$ (plus SPAM) as a function of $N$. As $N$ increases, $w_G$ becomes negligible, indicating that non-Markovian errors are suppressed under RC.}
    \label{fig:figure5}
\end{figure*}

\noindent
In the context of quantum computing, in particular for characterization and benchmarking, an error process is typically considered to be non-Markovian if it cannot be modeled by a process matrix. More precisely, non-Markovian errors are present if each $n$-qubit cycle (or layer) of gates cannot be modeled by a fixed, context-independent $n$-qubit process matrix. Various common sources of non-Markovianity in NISQ systems include leakage out of the computation basis states \cite{ghosh2013understanding, wallman2016robust, chen2016measuring, wood2018quantification, hayes2020eliminating, babu2021state}, unwanted entangling interactions (e.g., static $ZZ$ coupling) with qubits outside of the studied system, drift in qubit parameters \cite{proctor2020detecting} (e.g., stochastic fluctuations in transition frequencies), unwanted coupling to environmental systems with memory beyond the timescale of a cycle (e.g., two-level fluctuators and nonequilibrium quasiparticles \cite{serniak2018hot, de2020two, wilen2021correlated, berlin2022changes}), and qubit heating \cite{webb2018resilient}; see Fig.~\ref{fig:figure5}a. Studying and suppressing non-Markovian errors is important for at least two reasons: they interfere with the quantification of Markovian errors, and their impacts on QEC are less well-understood.

GST is designed to reconstruct all possible Markovian errors on any cycle of quantum gates. So, when GST cannot fit the data, this implies that there are non-Markovian errors. While model violation $N_\sigma$ is useful for providing evidence of the existence of non-Markovian errors, to quantify the magnitude of such errors, we add a \emph{wildcard error model} \cite{blume2020wildcard} to each of our GST models. These wildcard error models assign a wildcard error rate $w_G \in [0,1]$ to each gate $G$, and a wildcard error rate to the SPAM $w_{\textrm{SPAM}} \in [0,1]$, which quantifies how much additional error on each operation is missing from the model (i.e., how much is required to make the model consistent with the data; see Methods for more details). By comparing $w_G$ to $\epsilon_\diamond$ for each gate, we are able to quantify whether Markovian or unmodeled non-Markovian errors dominate the error model \cite{blume2020wildcard, rudinger2021experimental}. If $w_G \ll \epsilon_\diamond$ for all gates, then Markovian errors dominate and non-Markovian errors are negligible. In this case, the model captures the majority of the errors in the gate (or, more precisely, all those errors that were revealed in this experiment), despite the fact that there is evidence that the model is incomplete. On the other hand, if $w_G \ge \epsilon_\diamond$, the non-Markovian errors dominate the Markovian errors, and the so the GST estimate is unreliable. 

Fig.~\ref{fig:figure5}b shows the wildcard error and the diamond norm error for all of the gates in our gate set and SPAM, for $N=0,10,100$ (we omit the $N=1$ data due to its systematic inconsistency with a Markovian error model; see Model Selection and Methods). Without RC ($N=0$), we observe diamond norm errors as large as $\epsilon_\diamond \approx 0.07$, with up to $0.01$ additional non-Markovian error. For $N=10$ randomizations, the wildcard errors are still significant, but they are much smaller in magnitude than $N=0$ \emph{and} they are a small fraction of the diamond norm error rates. By $N=100$ randomizations, the wildcard errors are negligible --- in absolute terms and as a fraction of $\epsilon_{\diamond}$ --- contributing at most 0.0012 additional error per gate. This indicates that the S model accurately captures almost all of the $N=100$ data (note that in this case the wildcard error quantifies the combined contribution of both non-Markovian errors and all non-S Markovian errors). Because $w_G \ll \epsilon_\diamond \; \forall \; G \in \mathbb{G} $ for all $N$, we consider all of our models trustworthy, even for $N=0$. We additionally note that RC significantly improves both the wildcard error and worst-case error rate for SPAM, which is the sum of the trace distance for the state-preparation and the diamond norm for the positive operator-valued measure. However, even in the $N=100$ case, the SPAM error rate remains around $\sim 4\%$, well above the $1\%$ surface code threshold; these errors are equally as important as gate errors in the context of fault-tolerant QEC, and will need to be dramatically improved for achievable QEC in the future.

The $N=100$ data was gathered over a period of over 40 hours, and no re-calibration of gates was performed during the experiment. Therefore, the negligible amount of unmodeled error speaks to the robustness of RC to the inevitable drift in gate and qubit parameters during this time period. While RC was not designed to specifically target non-Markovian errors, its apparent robustness to both non-Markovian errors and correlated errors is promising for future large-scale fault-tolerant applications.
\section*{Discussion} \label{sec:discussion}

\noindent
The field of quantum characterization, verification, and validation (QCVV) was developed in part to benchmark our progress toward fault-tolerant QEC. While full-scale fault-tolerant QEC is still many years away, contemporary quantum gates are rapidly approaching the necessary thresholds for many QEC codes, such as the surface code. Therefore, it is important to be able to accurately benchmark our progress toward this goal. While average error rates measured via randomized benchmarks are useful for tracking progress, they fall short of capturing the information required to determine whether all gate errors fall below a given FT threshold, for which the diamond norm is the relevant metric \cite{sanders2015bounding, kueng2016comparing}. In this work, we demonstrate that FT thresholds can in fact be captured by randomized benchmarks, but only if the error-corrected application is performed using a randomization method which tailors noise. In fact, utilizing artificial randomness to suppress coherent dynamics is not new, and has been previously discussed for disorder-assisted error correction methods \cite{wootton2011bringing, stark2011localization, bravyi2012disorder}. While our results were measured in the many-shot, many-randomization limit, they are still relevant for error correction, which operates in the single-shot, single-randomization limit, because results measured in latter limit sample from the same distribution that is estimated in the former limit; in other words, results measured in the single-shot, single-randomization limit are indistinguishable from those sampled from a distribution impacted by only stochastic noise.

The current approach also helps resolve a discrepancy regarding logical error rates in the Pauli twirling approximation (PTA). While some have argued that logical noise is well approximated by Pauli errors for large-distance codes \cite{bravyi2018correcting}, others have shown that the PTA is a dishonest approximation of the impact of coherent errors \cite{geller2013efficient, tomita2014low, katabarwa2015logical, puzzuoli2014tractable, gutierrez2015comparison} and breaks down for large numbers of qubits \cite{katabarwa2017dynamical}, and that despite the projective nature of QEC, coherent errors at the physical level can lead to coherent errors at the logical level \cite{fern2006generalized, gutierrez2016errors, darmawan2017tensor, greenbaum2017modeling}, which can increase logical error rates \cite{hakkaku2021sampling}. Our results show that Pauli twirling can efficiently be implemented at the physical level, which may even impact the kinds of errors that manifest at the logical level, potentially negating any need to \emph{approximate} the accuracy of Pauli twirling at the logical level. Moreover, while coherent errors may continue persist even at the logical level, recent studies suggest that the diamond norm of gates at the logical level is smaller than the diamond norm of gates at the physical level \cite{huang2019performance}, which bodes favorably for the future of QEC. An exploration of this topic would be a natural extension of this work, and would provide insight into the necessity and/or potential benefits of using randomization methods for QEC.

While rapid improvements in two-qubit gates on many hardware platforms engender optimism for FT, caution must be taken in the claims inferred from gate fidelities, as results which include estimates of the diamond norm suggest that many contemporary two-qubit gates fall short of FT thresholds \cite{dehollain2016optimization, hughes2020benchmarking, mkadzik2021precision, white2021performance, xue2022quantum}. Furthermore, while isolated single- and two-qubit gates may be approaching the necessary requirements for fault-tolerance, any gate(s) performed in parallel with other qubits are likely to be impacted by crosstalk-induced coherent errors, potentially causing the diamond norm to scale with $\sqrt{e_F}$. The figure of merit for determining whether low logical error rates can be achieved via QEC is the error rate of a cycle containing all active qubits in a register, not simply the error rate of isolated gates within the cycle \cite{fowler2014quantifying}. Thus, whether or not randomization methods such as RC will be effective at the scale of the large number of qubits needed for QEC is an open question. However, in theory the efficiency of noise tailoring via Pauli twirling does not depending on system size \cite{goss2023extending}. Therefore, future work could explore the efficacy of RC at saturating the lower bound of the diamond norm for larger cycles of simultaneous gate operations.
\newline
\section*{Acknowledgements} \label{sec:acknowledgements}
\noindent
This work was supported by the U.S.~Department of Energy, Office of Science, Office of Advanced Scientific Computing Research Quantum Testbed Program under Contract No.~DE-AC02-05CH11231 and the Quantum Testbed Pathfinder Program.

\section*{Competing interests.}
\noindent
The Authors declare no Competing Financial or Non-Financial Interests.

\section*{Author Contributions}
\noindent
A.H.~conceptualized and performed the experiments. K.R., S.S., T.P., and N.G.~assisted in the design of the experiments. A.H.~analyzed the data with the help of K.R., S.S., and T.P. J.M.K. fabricated the sample. R.K.N., D.I.S., and I.S.~supervised all experimental work. A.H.~and T.P.~wrote the manuscript with input from all coauthors.

A.H.~acknowledges fruitful discussions with Joel Wallman, Joseph Emerson, Robin Blume-Kohout, Samuele Ferracin, and Long Nguyen. T.P.~acknowledges helpful discussions with Kevin Young and Robin Blume-Kohout.

\section*{Correspondence.}
\noindent
All correspondence and requests for materials should be addressed to A.H.

\section*{Data Availability.}
\noindent
All data is available from the corresponding author upon reasonable request.

\section*{Code Availability.}
\noindent
All GST sequences were generated and analyzed using \texttt{pyGSTi} \cite{nielsen2019python, nielsen2020probing}, an open-source Python package. All CB and CER sequences were generated and analyzed using \texttt{True-Q} \cite{beale_stefanie_j_2020_3945250}, a proprietary software package.

\section*{Methods}\label{sec:methods}
\small

\paragraph*{\textbf{\textup{Experiment.}}}\label{sec:experiment}
\noindent
All experimental gate set tomography (GST) results were measured in a manner which normalizes shot statistics between circuits implemented with and without randomized compiling (RC). Bare GST sequences were measured $K=1000$ times. For GST sequences implemented with $N$ randomizations under RC, we fix $K=1000$ and measure each randomization $K/N$ times. By computing the union over the $N$ different logically-equivalent randomizations of each GST sequence, one obtains an equivalent statistical distribution for a single GST circuit measured $K$ times.
\newline

\paragraph*{\textbf{\textup{Qubit Parameters.}}}\label{sec:single_qubit_parameters}

\noindent
Table \ref{tab:single_qubit_parameters} lists the relevant qubit parameters for the two transmon qubits used in this work. Qubit frequencies and anharmonicities are measured using Ramsey spectroscopy. Relaxation ($T_1$) and coherence ($T_2^*$ and $T_2^\text{echo}$) times are extracted by fitting exponential decay curves to the excited state lifetime and Ramsey spectroscopy measurements (without and with an echo pulse), respectively. The reader is referred to Ref.~\cite{hashim2021randomized} for details of the sample \texttt{AQT@LBNL Trailblazer8-v5.c2} and the experimental setup of the device, as well as Refs.~\cite{mitchell2021hardware, rudinger2021experimental, hashim2021optimized, ferracin2022efficiently} for previous characterization of the qubits used in this experiment.
\newline

\begin{table}[ht] 
  \centering
  \resizebox{0.75\columnwidth}{!}{
  \begin{tabular}{|l||r|r|r|r|}
    \hline
    {} & Q5 & Q6 \\
    \hline
    \hline
    Qubit freq.~(GHz) & 5.331004 & 5.490952 \\
    Anharm. (MHz) & -275 & -271.35 \\
    $T_1$ ($\mu$s) & 62(5) & 52(4) \\
    $T_2^*$ ($\mu$s) & 37(6) & 36(6) \\
    $T_2^\text{echo}$ ($\mu$s) & 73(7) & 68(7) \\
    \hline
  \end{tabular}}
\caption{Single-qubit parameters}\label{tab:single_qubit_parameters}
\end{table}

\paragraph*{\textbf{\textup{Representations of Quantum Processes.}}}\label{sec:rep_quant_op}
\noindent
Various representations of quantum processes exist \cite{greenbaum2015introduction}. A common representation is the operator-sum, or Kraus, representation which maps any density matrix $\rho$ into
\begin{equation}
    \mathcal{E}(\rho) = \sum_i K_i \rho K_i^\dagger,
\end{equation}
where the set $\{K_i\}$ is a general class of operators known as Kraus operators. For a Pauli channel, 
\begin{equation}\label{eq:kraus_pauli}
    \mathcal{E}(\rho) = \sum_{P \in \mathbb{P}^{\otimes n}} c_P P \rho P^\dagger
\end{equation}
where the Kraus operators are $K = \sqrt{c_P}P$, with $c_P$ the probability that $P$ is applied (which is an error, except in the case of the identity Pauli), and $\mathbb{P}^{\otimes n} = \{I, X, Y, Z\}^{\otimes n}$ is the set of $4^{n}$ generalized Pauli operators.

Another useful representation of a quantum operator is the Pauli transfer matrix (PTM) representation, which is used throughout this work. The PTM representation can be defined by expanding any density matrix $\rho$ in the Pauli basis,
\begin{equation}
    \rho = \sum_{P \in \mathbb{P}^{\otimes n}} \alpha_P P,
\end{equation}
where $\alpha_P$ are the expansion coefficients. By vectorizing the expansion coefficients, we obtain a vectorized density matrix $\ket{\rho}\rangle = \begin{pmatrix} \alpha_{I^{\otimes n}} & ... & \alpha_{Z^{\otimes n}} \end{pmatrix}^T$. A linear completely-positive and trace-preserving (CPTP) quantum map $\rho \mapsto \rho' = \Lambda(\rho)$ in the PTM representation is completely defined by a $4^n \times 4^n$ matrix $\Lambda$, with values $\Lambda_{ij} = \Tr[P_i\mathcal{E}(P_j)]/d$ that can derived directly from the Kraus representation (Eq.~\ref{eq:kraus_pauli}) for a Hilbert space of dimension $d = 2^n$. All entries in the PTM are real and are bounded by $\Lambda_{ij} \in [-1, 1]$. PTMs have the useful property that the composite map (i.e.~the PTM of a quantum circuit) can be constructed by taking the matrix products of the individual maps (i.e.~the PTMs of the individual gates).

For a single qubit,
\begin{equation}
    \ket{\rho}\rangle = \begin{pmatrix}
                        \alpha_{I} \\
                        \alpha_{X} \\
                        \alpha_{Y} \\
                        \alpha_{Z}
                        \end{pmatrix}.
\end{equation}
The quantum map $\rho' = \Lambda(\rho)$ can be expressed in vector form, $\ket{\rho'}\rangle = \Lambda \ket{\rho}\rangle$, or more explicitly, 
\begin{equation}
    \begin{pmatrix}
        \alpha_{I}' \\
        \alpha_{X}' \\
        \alpha_{Y}' \\
        \alpha_{Z}
    \end{pmatrix} = 
    \begin{pmatrix}
        \Lambda_{II} & \Lambda_{IX} & \Lambda_{IY} & \Lambda_{IZ} \\
        \Lambda_{XI} & \Lambda_{XX} & \Lambda_{XY} & \Lambda_{XZ} \\
        \Lambda_{YI} & \Lambda_{YX} & \Lambda_{YY} & \Lambda_{YZ} \\
        \Lambda_{ZI} & \Lambda_{ZX} & \Lambda_{ZY} & \Lambda_{ZZ}
    \end{pmatrix}
    \begin{pmatrix}
        \alpha_{I} \\
        \alpha_{X} \\
        \alpha_{Y} \\
        \alpha_{Z}
    \end{pmatrix}.
\end{equation}
As shown in Fig.~\ref{fig:figure1}a, some important properties of a process can be easily extracted from the components of its PTM. The PTM can be divided into four blocks: the upper left-hand corner represents trace-preservation, with $\Lambda_{II} = 1$ if a process is trace-preserving (TP). This constraint can be succinctly summarized by stating that a process is TP if $\Lambda_{0j} = \delta_{0j}$ (i.e.~the first row of the PTM is $[1, 0, ..., 0]$). The lower right-hand block is the unital block, which captures processes such as stochastic Pauli noise and unitary errors. A unital process is one that maps the identity operation to the identity operation $\Lambda(\mathbb{I}) = \mathbb{I}$ (it cannot purify a mixed state). The row above the unital block captures state-dependent leakage, represented by $\Lambda_{IP} \neq 0$ for $P \in \{X, Y, Z\}$; leakage is therefore not TP. The column to the left of the unital block is the non-unital block, which captures processes such as spontaneous emission (i.e.~$T_1$ decay) or amplitude damping. This constraint can be summarized by stating that a process is unital if $\Lambda_{i0} = \delta_{i0}$ (i.e.~the first column of the PTM is $[1, 0, ..., 0]^T$). Finally, the diagonal of $\Lambda$ represents state-preservation, with $\Lambda_{PP} = 1$ ($< 1$) for processes which (do not) preserve the Pauli channel $P \; \forall \; \{I, X, Y, Z\}$.

The process fidelity of a map in the PTM representation is the weighted average of the diagonal components,
\begin{equation}
    F_\Lambda = \frac{1}{4^n} \sum_{P \in \mathbb{P}^{\otimes n}} \Lambda_{PP}.
\end{equation}
Therefore, the process infidelity is given by $e_F = 1 - F_\Lambda$.
\newline

\paragraph*{\textbf{\textup{Gate Set Tomography.}}}\label{sec:gst_methods}
\noindent
In GST, the quality of the model is quantified by computing the log-likelihood ratio $\lambda$ of the likelihood $\mathcal{L}$ of the GST model with the likelihood $\mathcal{L}_\text{max}$ of the ``maximal model'',
\begin{equation}
    \lambda = -2 \ln{\left( \frac{\mathcal{L}}{\mathcal{L}_\text{max}} \right)}.
\end{equation}
The maximal model is the one in which each independent measurement outcome in the data set is assigned a distinct probability equal to the observed frequencies. Wilks' theorem \cite{wilks1938large} states that if $\mathcal{L}$ and $\mathcal{L}_\text{max}$ are both valid models, then the log-likelihood ratio is a $\chi^2_k$ random variable, where $k = N_\text{max} - N_p$ is the difference in the number of free parameters between the maximal and GST models. If $\lambda$ is not consistent with $\chi^2_k$ distribution (i.e.~it does not lie within the interval $[k - \sqrt{2k}, k + \sqrt{2k}]$, with mean $k$ and standard deviation $\sqrt{k}$), then this indicates that the GST data are inconsistent with the GST model.

We quantify the model violation, or ``goodness of fit,'' by the number of standard deviations that $\lambda$ is from the expected mean $k$,
\begin{equation}
    N_\sigma = \frac{\lambda - k}{\sqrt{2k}}.
\end{equation}
If $N_\sigma \le 1$, then the GST model can be considered trustworthy and faithfully captures the behavior of the device. GST makes an assumption of Markovianity (i.e.~any Markovian process can --- by definition --- be captured in a generalized model based on process matrices), therefore large $N_\sigma$ indicates the presence of non-Markovian errors. However, $N_\sigma$ does not quantify the magnitude of these non-Markovian errors. Because $N_\sigma$ will grow linearly with the number of shots (and will typically increase with the depth of the GST circuits), a large $N_\sigma$ can be observed even if the underlying non-Markovian errors are small in magnitude. Therefore, large $N_\sigma$ simply indicates that some non-Markovianity is present with high statistical certainty. 

While $N_\sigma$ provides statistical evidence of non-Markovianity, a wildcard error model \cite{blume2020wildcard} can capture the magnitude of these errors. The wildcard error rate $w_G \in [0,1]$ quantifies the unmodeled error per logic gate operation. A wildcard error can also be assigned for a circuit $C$ containing gates by summing over the wildcard error rates for all gates $G \in C$: $w_C = \sum_{G \in C}w_G$. The wildcard model is chosen to be minimal, such that assigning $w_G$ to a gate $G$ is just sufficient to make the model's predictions consistent with the observed data. This is enforced by requiring that the total variation distance (TVD)
\begin{equation}
    D_\text{TV}(p, q) = \frac{1}{2} ||p - q||_1
\end{equation}
between the observed probability distribution $p_C$ and the wildcard-augmented probability distribution $q_C$ be bounded by the total wildcard error for circuit $C$,
\begin{equation}\label{eq:tvd_wildcard}
    D_\text{TV}(p_C, q_C) \le w_C.
\end{equation}
The wildcard-augmented model is therefore not unique, as $q_C$ can be chosen from any distribution that satisfies Eq.~\ref{eq:tvd_wildcard}. Because the wildcard error quantifies the magnitude of unmodeled error per gate, and because unmodeled errors are often attributed to non-Markovianity in the system, the per-gate wildcard error budget is a good estimate of the magnitude of non-Markovian errors impacting the gate. The TVD is a useful metric for quantifying the magnitude of unmodeled errors because it captures the rate at which measurement outcomes are incorrectly predicted by a model. Because the TVD is upper-bounded by the diamond norm \cite{kitaev1997quantum}, one can compare the wildcard error $w_G$ to the diamond error $\epsilon_\diamond$ for any gate $G$ to inform whether unmodeled errors in the GST estimate are dominant or negligible, and thus whether the GST model can be trusted. By extension, comparing $w_G$ to $\epsilon_\diamond$ quantifies whether Markovian or non-Markovian errors dominate the total error.
\newline


\paragraph*{\textbf{\textup{Randomization Compiling: single-randomization limit.}}}\label{sec:rc_srl}
\noindent
To understand how RC suppresses the coherent accumulation of unitary errors between cycles of gates, consider the simple example of applying many $R_x(2\pi)$ rotations to a qubit in the ground state, but each time the qubit over-rotates by a small angle $\theta$. The resulting state of the qubit after $M$ rotations is
\begin{equation}
    \ket{\psi} = \prod^M  e^{-i\theta\sigma_x} \ket{0} = \cos{(M\theta)}\ket{0} - i\sin{(M\theta)}\ket{1}.
\end{equation}
The fidelity of this state with respect to $|0\rangle$ is $\mathcal{F} = \vert \langle 0 | \psi\rangle \vert^2 = \cos^2{(M\theta)} \approx 1 - (M\theta)^2$, thus the infidelity $r = 1 - \mathcal{F} \approx (M\theta)^2$. Therefore, the infidelity scales quadratically in both the over-rotation angle $\theta$ and the number of rotations $M$. Under RC, the trajectory from the initial state to the final state is randomized, thus ensuring that coherent errors will not grow quadratically between gates. While exact quadratic growth is highly unlikely for longer-depth multi-qubit circuits due to the complex dynamics of crosstalk, and because coherent errors can act in any direction (not just along the axis of rotation), coherent errors can still accumulate in an adversarial fashion and grow faster than average error rates, especially in structured quantum circuits \cite{proctor2021measuring}.
\newline

\paragraph*{\textbf{\textup{Randomization Compiling: many-randomization limit.}}}\label{sec:rc_mrl}
\noindent
Regardless of the rate at which coherent errors accumulate between cycles of gates, they can still impact each computational gate $G$ in a circuit. We can model this process as an ideal gate $G_0$ followed by an unwanted unitary operator $G = U(\mathbf{\hat{n}}, \theta) G_0$, where $U(\mathbf{\hat{n}}, \theta) = e^{-i\theta\mathbf{\hat{n} \cdot \boldsymbol\sigma}/2}$, $\mathbf{\hat{n}}$ is the axis of rotation, $\boldsymbol\sigma$ the Pauli vector, and $\theta$ is the rotation angle relative to the intended target state. For simplicity, consider a unitary error about the $x$-axis for a single qubit,
\begin{align}
    U(x, \theta) &= \exp\left( - i \frac{\theta}{2} \sigma_x \right) \nonumber \\
    &= \left( \begin{array}{cc} \cos(\theta / 2) & - i \sin(\theta / 2) \\  
                              i \sin(\theta / 2) & \cos(\theta / 2) \end{array} \right).
\end{align}
In the PTM representation (see Fig.~\ref{fig:figure1}a), this coherent error takes the following form, 
\begin{equation}
   \Lambda = \left(
        \begin{array}{cccc}
            1 & 0 & 0 & 0 \\
            0 & 1 & 0 & 0 \\
            0 & 0 & \cos(\theta) & - \sin(\theta) \\
            0 & 0 & \sin(\theta) & \cos(\theta) \\
        \end{array}
        \right).
\end{equation}
For small $\theta$, the diagonal components of $\Lambda$ scale as $\cos(\theta) \approx 1 - \tfrac{1}{2}\theta^2$, and the off-diagonal terms scale as $\sin(\theta) \approx \theta$. While the infidelity of the diagonal terms is $e_F \approx \theta^2$, we see that the off-diagonal terms are quadratically larger, with $\sqrt{e_F} \approx \theta$. While not all error metrics are sensitive to the off-diagonal terms in an error process (e.g.~fidelity-based measures), norm-based error metrics such as the diamond norm generally are sensitive to such terms, setting the $\sim \sqrt{e_F}$ upper bound of Eq.~\ref{eq:bounds}.

Under RC in the many-randomization limit, all off-diagonal terms in the error process are completely suppressed in the limit that $N \longrightarrow \infty$. To understand how this occurs, consider Pauli twirling $\Lambda$, i.e.~$P \Lambda P^\dagger$ for any $P \in \{I, X, Y, Z \}$, where $P$ represents the Pauli superoperator. Under Pauli conjugation, the signs of the off-diagonal terms remain the same (are reversed) if $P$ commutes (does not commute) with $\Lambda$. Thus, the off-diagonal terms change sign with a 50\% probability upon conjugation with a randomly-selected Pauli. When averaging over $N$ randomizations, the magnitude of the off-diagonal terms scale as $\theta/\sqrt{N}$, reminiscent  of a random walk, and thus vanish as $N \longrightarrow \infty$ or if by luck the correct Paulis were sampled which average to zero. While the ``noise tailoring'' property of RC rests on assumption that the noise impacting the easy gates is gate-independent, Wallman \textit{et.~al.~}\cite{wallman2016noise} prove that RC is robust to small gate-dependent errors, which are inevitable in modern-day experiments.
\newline

\paragraph*{\textbf{\textup{Method for Performing Randomized Compiling on GST Sequences.}}}\label{sec:rc_methods}
\noindent
In order to preserve the circuit depth of GST sequences under RC, randomly sampled single-qubit Paulis and their correction gates are inserted between every layer. For circuits only containing single-qubit gates, the random Paulis are compiled into the previous layer and the correction gates are compiled into the subsequent layer. For circuits containing two-qubit gates, the correction gates are commuted through the two-qubit gate before being compiled into the subsequent layer.

To highlight this method, consider a circuit $\mathcal{C}$ containing $N$ layers $L$ of single-qubit gates:
\begin{equation}
    \mathcal{C} = L_N L_{N-1}  ... L_3 L_2 L_1.
\end{equation}
Under RC, a single randomized circuit takes the following form
\begin{equation}
    \mathcal{C} = P_{N} L_N P_{N-1}^\dagger P_{N-1} L_{N-1} ... L_3 P_2^\dagger P_2 L_2 P_1^\dagger P_1 L_1,
\end{equation}
where $P_{N}^\dagger$ is omitted in the circuit but taken into account in the final ideal measurement results. The compiled circuit is
\begin{equation}
    \mathcal{C} = \tilde{L}_N \tilde{L}_{N-1} ... \tilde{L}_3 \tilde{L}_2\tilde{L}_1,
\end{equation}
where the $k$th layer $\tilde{L}_k = P_k L_k P_{k-1}^\dagger$ (except for the first layer, which does not contain a correction gate). For circuits containing two-qubit gates in layer $k-1$, the $k$th layer becomes $\tilde{L}_k = P_k L_k P_{k-2}^c$, where $P_{k-2}^c = L_{k-1} P_{k-2}^\dagger L_{k-1}^\dagger$. This method therefore randomizes all layers of single-qubit gates, while also maintaining the original circuit depth, and was developed within the \texttt{pyGSTi} framework specifically for the purpose of randomizing GST and related benchmarking circuits.

As highlighted in the main text, we observe larger model violation for a single-randomization under RC than for no randomizations. This is expected behavior for $N=1$ even when the physical gates are Markovian, which we illustrate with the following example: consider two single-qubit circuits, $\mathcal{C}_1 = G_I G_I$ and $\mathcal{C}_2 = G_I G_I G_I G_I$, and consider a physical gate set $\{G_I, G_X, G_Y, G_Z\}$, with $G_X = X_\pi$, etc. Consider a simple error model where each gate is followed by a small coherent $X_\theta$ rotation error. When we implement $\mathcal{C}_1$ with a single randomization under RC, we actually implement a circuit that should perform the identity rotation, but the gates have been randomized according to the method outlined above. Let's assume our randomized circuit $\tilde{\mathcal{C}}_1$ is the same as the original circuit, $\tilde{\mathcal{C}}_1 = \mathcal{C}_1 = G_I G_I$. If we initialize our qubit in the ground state, we will observe a small coherent over-rotation error by $2\theta$, which will result in some $\theta$-dependent probability of not measuring 0. Similarly, when we run $\mathcal{\mathcal{C}}_2$ we will actually actually implement one of the many length-4 combinations of Pauli gates that produces the identity rotation. For example, suppose we sampled $\tilde{\mathcal{C}}_2 = G_Z G_Z G_Z G_Z$; this combination of gates would echo away the $X$ rotation error, and we would measure 0 with probability 1. Therefore, altogether we observe $\Pr(0|\mathcal{C}_1) < 1$ and $\Pr(0|\mathcal{C}_2) = 1$. This is inconsistent with every possible process matrix for $G_I$ (or at least every process matrix that is close to the target identity matrix), because repeating an identity gate amplifies all of its error parameters, but does not echo away errors. Finally, note that this argument is predicated on the assumption that we measure each circuit many times; if we measure each circuit only once, the results will not be inconsistent, but also will not be very informative.
\newline

\begin{figure}[h]
    \centering
    \includegraphics[width=\columnwidth]{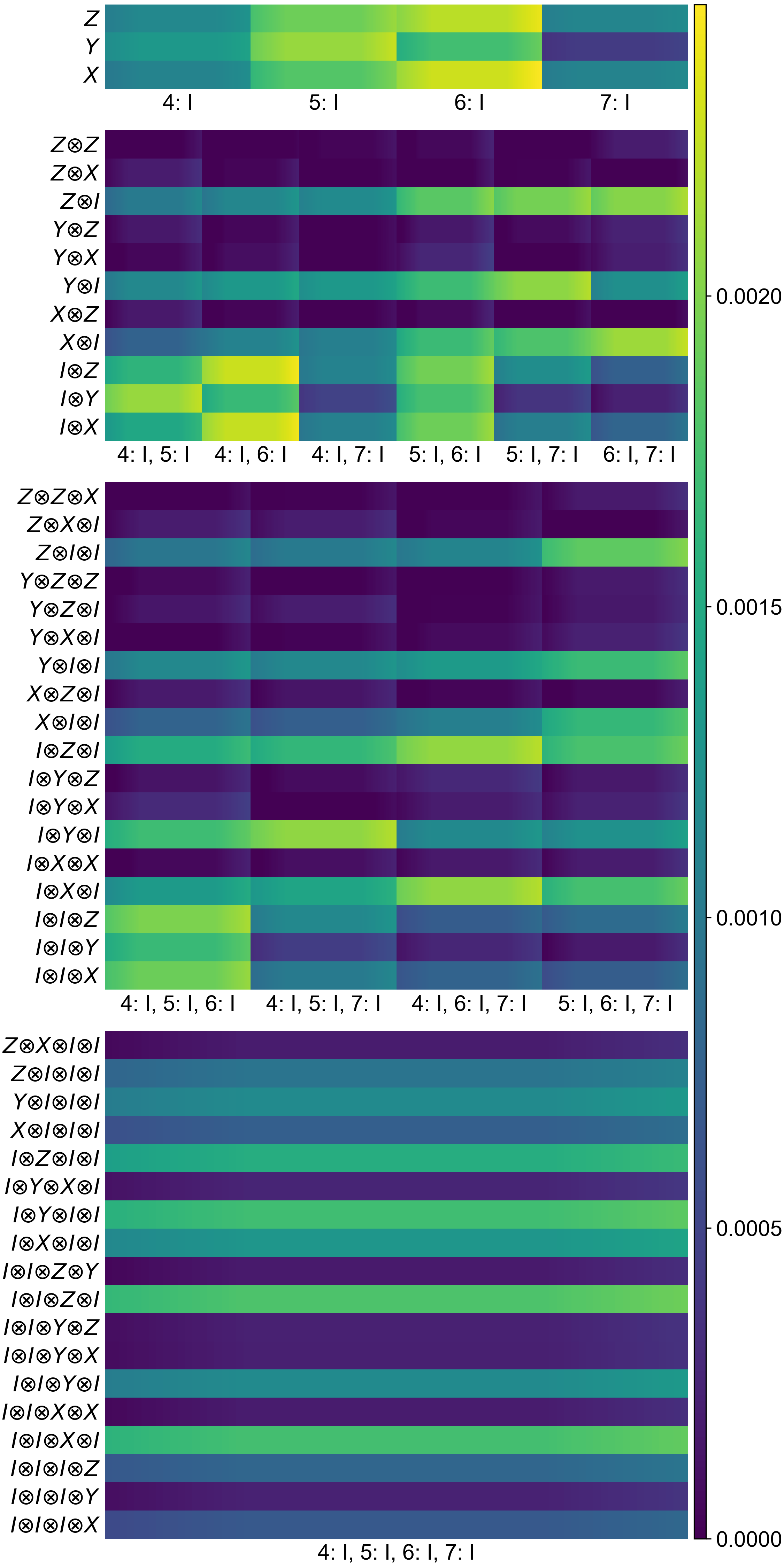}
    \caption{Cycle error reconstruction of the identity cycle on four qubits $\{I_4 \otimes I_5 \otimes I_6 \otimes I_7\}$. Weight-1 errors dominate the error map; all weight-$k \ge 2$ errors are negligible in comparison. Rows in which all errors fall below 12\% of the maximum value have been omitted for visual clarity.}
    \label{fig:knr4}
\end{figure}

\paragraph*{\textbf{\textup{Cycle Error Reconstruction.}}}\label{sec:cer_methods}
\noindent
In Fig.~\ref{fig:knr4}, we plot the results of CER applied to the identity cycle $\{I_4 \otimes I_5 \otimes I_6 \otimes I_7\}$ across all four qubits on the quantum processor. We measure all $4^4 = 256$ Pauli channels, but omit values which fall below 12\% of the maximum value for visual clarity. We observe no significant contribution from weight-$k \ge 2$ in the plot, demonstrating that Pauli twirling can break larger-scale correlations across an entire quantum processor. Furthermore, Refs.~\cite{hashim2021randomized, ferracin2022efficiently} present similar results for multi-qubit cycles containing two-qubit gates, and the conclusions were the same: the most dominant errors under Pauli twirling are single-body errors (i.e.~weight-1 errors for single-qubit gates, weight-1 or weight-2 errors for two-qubit gates, etc.). Moreover, because two-body errors are the sum of all errors that act non-trivially on the corresponding two bodies, irrespective of their action on other qubits, the fact that two-body errors are negligible indicates that higher-body errors are also negligible.

\clearpage
\bibliographystyle{naturemag}
\bibliography{bibliography}

\begin{thebibliography}{100}
\expandafter\ifx\csname url\endcsname\relax
  \def\url#1{\texttt{#1}}\fi
\expandafter\ifx\csname urlprefix\endcsname\relax\def\urlprefix{URL }\fi
\providecommand{\bibinfo}[2]{#2}
\providecommand{\eprint}[2][]{\url{#2}}

\bibitem{preskill2018quantum}
\bibinfo{author}{Preskill, J.}
\newblock \bibinfo{title}{Quantum computing in the nisq era and beyond}.
\newblock \emph{\bibinfo{journal}{Quantum}} \textbf{\bibinfo{volume}{2}},
  \bibinfo{pages}{79} (\bibinfo{year}{2018}).

\bibitem{shor1999polynomial}
\bibinfo{author}{Shor, P.~W.}
\newblock \bibinfo{title}{Polynomial-time algorithms for prime factorization
  and discrete logarithms on a quantum computer}.
\newblock \emph{\bibinfo{journal}{SIAM review}} \textbf{\bibinfo{volume}{41}},
  \bibinfo{pages}{303--332} (\bibinfo{year}{1999}).

\bibitem{shor1995scheme}
\bibinfo{author}{Shor, P.~W.}
\newblock \bibinfo{title}{Scheme for reducing decoherence in quantum computer
  memory}.
\newblock \emph{\bibinfo{journal}{Phys. Rev. A}} \textbf{\bibinfo{volume}{52}},
  \bibinfo{pages}{R2493} (\bibinfo{year}{1995}).

\bibitem{gottesman1996class}
\bibinfo{author}{Gottesman, D.}
\newblock \bibinfo{title}{Class of quantum error-correcting codes saturating
  the quantum hamming bound}.
\newblock \emph{\bibinfo{journal}{Phys. Rev. A}} \textbf{\bibinfo{volume}{54}},
  \bibinfo{pages}{1862} (\bibinfo{year}{1996}).

\bibitem{steane1996multiple}
\bibinfo{author}{Steane, A.}
\newblock \bibinfo{title}{Multiple-particle interference and quantum error
  correction}.
\newblock \emph{\bibinfo{journal}{Proc. Math. Phys. Eng. Sci. P ROY SOC A-MATH
  PHY}} \textbf{\bibinfo{volume}{452}}, \bibinfo{pages}{2551--2577}
  (\bibinfo{year}{1996}).

\bibitem{steane1998introduction}
\bibinfo{author}{Steane, A.}
\newblock \bibinfo{title}{Introduction to quantum error correction}.
\newblock \emph{\bibinfo{journal}{Philos. Trans. Royal Soc. A PHILOS T R SOC
  A}} \textbf{\bibinfo{volume}{356}}, \bibinfo{pages}{1739--1758}
  (\bibinfo{year}{1998}).

\bibitem{calderbank1998quantum}
\bibinfo{author}{Calderbank, A.~R.}, \bibinfo{author}{Rains, E.~M.},
  \bibinfo{author}{Shor, P.} \& \bibinfo{author}{Sloane, N.~J.}
\newblock \bibinfo{title}{Quantum error correction via codes over gf (4)}.
\newblock \emph{\bibinfo{journal}{IEEE Trans. Inf. Theory}}
  \textbf{\bibinfo{volume}{44}}, \bibinfo{pages}{1369--1387}
  (\bibinfo{year}{1998}).

\bibitem{shor1996fault}
\bibinfo{author}{Shor, P.~W.}
\newblock \bibinfo{title}{Fault-tolerant quantum computation}.
\newblock In \emph{\bibinfo{booktitle}{Proceedings of 37th Conference on
  Foundations of Computer Science}}, \bibinfo{pages}{56--65}
  (\bibinfo{organization}{IEEE}, \bibinfo{year}{1996}).

\bibitem{knill1996concatenated}
\bibinfo{author}{Knill, E.} \& \bibinfo{author}{Laflamme, R.}
\newblock \bibinfo{title}{Concatenated quantum codes}.
\newblock \emph{\bibinfo{journal}{Preprint at quant-ph/9608012}}
  (\bibinfo{year}{1996}).

\bibitem{knill1998resilient}
\bibinfo{author}{Knill, E.}, \bibinfo{author}{Laflamme, R.} \&
  \bibinfo{author}{Zurek, W.~H.}
\newblock \bibinfo{title}{Resilient quantum computation}.
\newblock \emph{\bibinfo{journal}{Science}} \textbf{\bibinfo{volume}{279}},
  \bibinfo{pages}{342--345} (\bibinfo{year}{1998}).

\bibitem{preskill1998fault}
\bibinfo{author}{Preskill, J.}
\newblock \bibinfo{title}{Fault-tolerant quantum computation}.
\newblock In \emph{\bibinfo{booktitle}{Introduction to quantum computation and
  information}}, \bibinfo{pages}{213--269} (\bibinfo{publisher}{World
  Scientific}, \bibinfo{year}{1998}).

\bibitem{kitaev2003fault}
\bibinfo{author}{Kitaev, A.~Y.}
\newblock \bibinfo{title}{Fault-tolerant quantum computation by anyons}.
\newblock \emph{\bibinfo{journal}{Ann. Phys.}} \textbf{\bibinfo{volume}{303}},
  \bibinfo{pages}{2--30} (\bibinfo{year}{2003}).

\bibitem{aharonov2008fault}
\bibinfo{author}{Aharonov, D.} \& \bibinfo{author}{Ben-Or, M.}
\newblock \bibinfo{title}{Fault-tolerant quantum computation with constant
  error rate}.
\newblock \emph{\bibinfo{journal}{SIAM J. Sci. Comput.}}
  (\bibinfo{year}{2008}).

\bibitem{aliferis2005quantum}
\bibinfo{author}{Aliferis, P.}, \bibinfo{author}{Gottesman, D.} \&
  \bibinfo{author}{Preskill, J.}
\newblock \bibinfo{title}{Quantum accuracy threshold for concatenated
  distance-3 codes}.
\newblock \emph{\bibinfo{journal}{Quantum Info. Comput.}}
  \textbf{\bibinfo{volume}{6}}, \bibinfo{pages}{97–165}
  (\bibinfo{year}{2006}).

\bibitem{aliferis2007subsystem}
\bibinfo{author}{Aliferis, P.} \& \bibinfo{author}{Cross, A.~W.}
\newblock \bibinfo{title}{Subsystem fault tolerance with the bacon-shor code}.
\newblock \emph{\bibinfo{journal}{Phys. Rev. Lett.}}
  \textbf{\bibinfo{volume}{98}}, \bibinfo{pages}{220502}
  (\bibinfo{year}{2007}).

\bibitem{aliferis2007accuracy}
\bibinfo{author}{Aliferis, P.}, \bibinfo{author}{Gottesman, D.} \&
  \bibinfo{author}{Preskill, J.}
\newblock \bibinfo{title}{Accuracy threshold for postselected quantum
  computation}.
\newblock \emph{\bibinfo{journal}{Quantum Info. Comput.}}
  \textbf{\bibinfo{volume}{8}}, \bibinfo{pages}{181–244}
  (\bibinfo{year}{2008}).

\bibitem{aliferis2009fibonacci}
\bibinfo{author}{Aliferis, P.} \& \bibinfo{author}{Preskill, J.}
\newblock \bibinfo{title}{Fibonacci scheme for fault-tolerant quantum
  computation}.
\newblock \emph{\bibinfo{journal}{Phys. Rev. A}} \textbf{\bibinfo{volume}{79}},
  \bibinfo{pages}{012332} (\bibinfo{year}{2009}).

\bibitem{chamberland2016thresholds}
\bibinfo{author}{Chamberland, C.}, \bibinfo{author}{Jochym-O’Connor, T.} \&
  \bibinfo{author}{Laflamme, R.}
\newblock \bibinfo{title}{Thresholds for universal concatenated quantum codes}.
\newblock \emph{\bibinfo{journal}{Phys. Rev. Lett.}}
  \textbf{\bibinfo{volume}{117}}, \bibinfo{pages}{010501}
  (\bibinfo{year}{2016}).

\bibitem{knill2005quantum}
\bibinfo{author}{Knill, E.}
\newblock \bibinfo{title}{Quantum computing with realistically noisy devices}.
\newblock \emph{\bibinfo{journal}{Nature}} \textbf{\bibinfo{volume}{434}},
  \bibinfo{pages}{39--44} (\bibinfo{year}{2005}).

\bibitem{aliferis2009fault}
\bibinfo{author}{Aliferis, P.} \emph{et~al.}
\newblock \bibinfo{title}{Fault-tolerant computing with biased-noise
  superconducting qubits: a case study}.
\newblock \emph{\bibinfo{journal}{New J. Phys.}} \textbf{\bibinfo{volume}{11}},
  \bibinfo{pages}{013061} (\bibinfo{year}{2009}).

\bibitem{duclos2010fast}
\bibinfo{author}{Duclos-Cianci, G.} \& \bibinfo{author}{Poulin, D.}
\newblock \bibinfo{title}{Fast decoders for topological quantum codes}.
\newblock \emph{\bibinfo{journal}{Phys. Rev. Lett.}}
  \textbf{\bibinfo{volume}{104}}, \bibinfo{pages}{050504}
  (\bibinfo{year}{2010}).

\bibitem{wang2011surface}
\bibinfo{author}{Wang, D.~S.}, \bibinfo{author}{Fowler, A.~G.} \&
  \bibinfo{author}{Hollenberg, L.~C.}
\newblock \bibinfo{title}{Surface code quantum computing with error rates over
  1\%}.
\newblock \emph{\bibinfo{journal}{Phys. Rev. A}} \textbf{\bibinfo{volume}{83}},
  \bibinfo{pages}{020302} (\bibinfo{year}{2011}).

\bibitem{bombin2012strong}
\bibinfo{author}{Bombin, H.}, \bibinfo{author}{Andrist, R.~S.},
  \bibinfo{author}{Ohzeki, M.}, \bibinfo{author}{Katzgraber, H.~G.} \&
  \bibinfo{author}{Martin-Delgado, M.~A.}
\newblock \bibinfo{title}{Strong resilience of topological codes to
  depolarization}.
\newblock \emph{\bibinfo{journal}{Phys. Rev. X}} \textbf{\bibinfo{volume}{2}},
  \bibinfo{pages}{021004} (\bibinfo{year}{2012}).

\bibitem{wootton2012high}
\bibinfo{author}{Wootton, J.~R.} \& \bibinfo{author}{Loss, D.}
\newblock \bibinfo{title}{High threshold error correction for the surface
  code}.
\newblock \emph{\bibinfo{journal}{Phys. Rev. Lett.}}
  \textbf{\bibinfo{volume}{109}}, \bibinfo{pages}{160503}
  (\bibinfo{year}{2012}).

\bibitem{stephens2014fault}
\bibinfo{author}{Stephens, A.~M.}
\newblock \bibinfo{title}{Fault-tolerant thresholds for quantum error
  correction with the surface code}.
\newblock \emph{\bibinfo{journal}{Phys. Rev. A}} \textbf{\bibinfo{volume}{89}},
  \bibinfo{pages}{022321} (\bibinfo{year}{2014}).

\bibitem{auger2017fault}
\bibinfo{author}{Auger, J.~M.}, \bibinfo{author}{Anwar, H.},
  \bibinfo{author}{Gimeno-Segovia, M.}, \bibinfo{author}{Stace, T.~M.} \&
  \bibinfo{author}{Browne, D.~E.}
\newblock \bibinfo{title}{Fault-tolerance thresholds for the surface code with
  fabrication errors}.
\newblock \emph{\bibinfo{journal}{Phys. Rev. A}} \textbf{\bibinfo{volume}{96}},
  \bibinfo{pages}{042316} (\bibinfo{year}{2017}).

\bibitem{tuckett2020fault}
\bibinfo{author}{Tuckett, D.~K.}, \bibinfo{author}{Bartlett, S.~D.},
  \bibinfo{author}{Flammia, S.~T.} \& \bibinfo{author}{Brown, B.~J.}
\newblock \bibinfo{title}{Fault-tolerant thresholds for the surface code in
  excess of 5\% under biased noise}.
\newblock \emph{\bibinfo{journal}{Phys. Rev. Lett.}}
  \textbf{\bibinfo{volume}{124}}, \bibinfo{pages}{130501}
  (\bibinfo{year}{2020}).

\bibitem{barends2014superconducting}
\bibinfo{author}{Barends, R.} \emph{et~al.}
\newblock \bibinfo{title}{Superconducting quantum circuits at the surface code
  threshold for fault tolerance}.
\newblock \emph{\bibinfo{journal}{Nature}} \textbf{\bibinfo{volume}{508}},
  \bibinfo{pages}{500--503} (\bibinfo{year}{2014}).

\bibitem{rong2015experimental}
\bibinfo{author}{Rong, X.} \emph{et~al.}
\newblock \bibinfo{title}{Experimental fault-tolerant universal quantum gates
  with solid-state spins under ambient conditions}.
\newblock \emph{\bibinfo{journal}{Nat. Commun.}} \textbf{\bibinfo{volume}{6}},
  \bibinfo{pages}{1--7} (\bibinfo{year}{2015}).

\bibitem{xue2022quantum}
\bibinfo{author}{Xue, X.} \emph{et~al.}
\newblock \bibinfo{title}{Quantum logic with spin qubits crossing the surface
  code threshold}.
\newblock \emph{\bibinfo{journal}{Nature}} \textbf{\bibinfo{volume}{601}},
  \bibinfo{pages}{343--347} (\bibinfo{year}{2022}).

\bibitem{emerson2005scalable}
\bibinfo{author}{Emerson, J.}, \bibinfo{author}{Alicki, R.} \&
  \bibinfo{author}{{\.Z}yczkowski, K.}
\newblock \bibinfo{title}{Scalable noise estimation with random unitary
  operators}.
\newblock \emph{\bibinfo{journal}{J. opt., B Quantum semiclass. opt.}}
  \textbf{\bibinfo{volume}{7}}, \bibinfo{pages}{S347} (\bibinfo{year}{2005}).

\bibitem{knill2008randomized}
\bibinfo{author}{Knill, E.} \emph{et~al.}
\newblock \bibinfo{title}{Randomized benchmarking of quantum gates}.
\newblock \emph{\bibinfo{journal}{Phys. Rev. A}} \textbf{\bibinfo{volume}{77}},
  \bibinfo{pages}{012307} (\bibinfo{year}{2008}).

\bibitem{dankert2009exact}
\bibinfo{author}{Dankert, C.}, \bibinfo{author}{Cleve, R.},
  \bibinfo{author}{Emerson, J.} \& \bibinfo{author}{Livine, E.}
\newblock \bibinfo{title}{Exact and approximate unitary 2-designs and their
  application to fidelity estimation}.
\newblock \emph{\bibinfo{journal}{Phys. Rev. A}} \textbf{\bibinfo{volume}{80}},
  \bibinfo{pages}{012304} (\bibinfo{year}{2009}).

\bibitem{magesan2011scalable}
\bibinfo{author}{Magesan, E.}, \bibinfo{author}{Gambetta, J.~M.} \&
  \bibinfo{author}{Emerson, J.}
\newblock \bibinfo{title}{Scalable and robust randomized benchmarking of
  quantum processes}.
\newblock \emph{\bibinfo{journal}{Phys. Rev. Lett.}}
  \textbf{\bibinfo{volume}{106}}, \bibinfo{pages}{180504}
  (\bibinfo{year}{2011}).

\bibitem{magesan2012efficient}
\bibinfo{author}{Magesan, E.} \emph{et~al.}
\newblock \bibinfo{title}{Efficient measurement of quantum gate error by
  interleaved randomized benchmarking}.
\newblock \emph{\bibinfo{journal}{Phys. Rev. Lett.}}
  \textbf{\bibinfo{volume}{109}}, \bibinfo{pages}{080505}
  (\bibinfo{year}{2012}).

\bibitem{kitaev1997quantum}
\bibinfo{author}{Kitaev, A.~Y.}
\newblock \bibinfo{title}{Quantum computations: algorithms and error
  correction}.
\newblock \emph{\bibinfo{journal}{Uspekhi Matematicheskikh Nauk}}
  \textbf{\bibinfo{volume}{52}}, \bibinfo{pages}{53--112}
  (\bibinfo{year}{1997}).

\bibitem{merkel2013self}
\bibinfo{author}{Merkel, S.~T.} \emph{et~al.}
\newblock \bibinfo{title}{Self-consistent quantum process tomography}.
\newblock \emph{\bibinfo{journal}{Phys. Rev. A}} \textbf{\bibinfo{volume}{87}},
  \bibinfo{pages}{062119} (\bibinfo{year}{2013}).

\bibitem{blume2013robust}
\bibinfo{author}{Blume-Kohout, R.} \emph{et~al.}
\newblock \bibinfo{title}{Robust, self-consistent, closed-form tomography of
  quantum logic gates on a trapped ion qubit}.
\newblock \emph{\bibinfo{journal}{Preprint at https://arxiv.org/abs/1310.4492}}
   (\bibinfo{year}{2013}).

\bibitem{greenbaum2015introduction}
\bibinfo{author}{Greenbaum, D.}
\newblock \bibinfo{title}{Introduction to quantum gate set tomography.}
\newblock \emph{\bibinfo{journal}{Preprint at
  https://arxiv.org/abs/1509.02921}}  (\bibinfo{year}{2015}).

\bibitem{blume2017demonstration}
\bibinfo{author}{Blume-Kohout, R.} \emph{et~al.}
\newblock \bibinfo{title}{Demonstration of qubit operations below a rigorous
  fault tolerance threshold with gate set tomography}.
\newblock \emph{\bibinfo{journal}{Nat. Commun.}} \textbf{\bibinfo{volume}{8}},
  \bibinfo{pages}{1--13} (\bibinfo{year}{2017}).

\bibitem{nielsen2021gate}
\bibinfo{author}{Nielsen, E.} \emph{et~al.}
\newblock \bibinfo{title}{Gate set tomography}.
\newblock \emph{\bibinfo{journal}{Quantum}} \textbf{\bibinfo{volume}{5}},
  \bibinfo{pages}{557} (\bibinfo{year}{2021}).

\bibitem{wallman2014randomized}
\bibinfo{author}{Wallman, J.~J.} \& \bibinfo{author}{Flammia, S.~T.}
\newblock \bibinfo{title}{Randomized benchmarking with confidence}.
\newblock \emph{\bibinfo{journal}{New J. Phys.}} \textbf{\bibinfo{volume}{16}},
  \bibinfo{pages}{103032} (\bibinfo{year}{2014}).

\bibitem{sanders2015bounding}
\bibinfo{author}{Sanders, Y.~R.}, \bibinfo{author}{Wallman, J.~J.} \&
  \bibinfo{author}{Sanders, B.~C.}
\newblock \bibinfo{title}{Bounding quantum gate error rate based on reported
  average fidelity}.
\newblock \emph{\bibinfo{journal}{New J. Phys.}} \textbf{\bibinfo{volume}{18}},
  \bibinfo{pages}{012002} (\bibinfo{year}{2015}).

\bibitem{wallman2015bounding}
\bibinfo{author}{Wallman, J.~J.}
\newblock \bibinfo{title}{Bounding experimental quantum error rates relative to
  fault-tolerant thresholds}.
\newblock \emph{\bibinfo{journal}{Preprint at
  https://arxiv.org/abs/1511.00727}}  (\bibinfo{year}{2015}).

\bibitem{kueng2016comparing}
\bibinfo{author}{Kueng, R.}, \bibinfo{author}{Long, D.~M.},
  \bibinfo{author}{Doherty, A.~C.} \& \bibinfo{author}{Flammia, S.~T.}
\newblock \bibinfo{title}{Comparing experiments to the fault-tolerance
  threshold}.
\newblock \emph{\bibinfo{journal}{Phys. Rev. Lett.}}
  \textbf{\bibinfo{volume}{117}}, \bibinfo{pages}{170502}
  (\bibinfo{year}{2016}).

\bibitem{sheldon2016characterizing}
\bibinfo{author}{Sheldon, S.} \emph{et~al.}
\newblock \bibinfo{title}{Characterizing errors on qubit operations via
  iterative randomized benchmarking}.
\newblock \emph{\bibinfo{journal}{Phys. Rev. A}} \textbf{\bibinfo{volume}{93}},
  \bibinfo{pages}{012301} (\bibinfo{year}{2016}).

\bibitem{arute2019quantum}
\bibinfo{author}{Arute, F.} \emph{et~al.}
\newblock \bibinfo{title}{Quantum supremacy using a programmable
  superconducting processor}.
\newblock \emph{\bibinfo{journal}{Nature}} \textbf{\bibinfo{volume}{574}},
  \bibinfo{pages}{505--510} (\bibinfo{year}{2019}).

\bibitem{wang2020high}
\bibinfo{author}{Wang, Y.} \emph{et~al.}
\newblock \bibinfo{title}{High-fidelity two-qubit gates using a
  microelectromechanical-system-based beam steering system for individual qubit
  addressing}.
\newblock \emph{\bibinfo{journal}{Phys. Rev. Lett.}}
  \textbf{\bibinfo{volume}{125}}, \bibinfo{pages}{150505}
  (\bibinfo{year}{2020}).

\bibitem{pino2021demonstration}
\bibinfo{author}{Pino, J.~M.} \emph{et~al.}
\newblock \bibinfo{title}{Demonstration of the trapped-ion quantum ccd computer
  architecture}.
\newblock \emph{\bibinfo{journal}{Nature}} \textbf{\bibinfo{volume}{592}},
  \bibinfo{pages}{209--213} (\bibinfo{year}{2021}).

\bibitem{mitchell2021hardware}
\bibinfo{author}{Mitchell, B.~K.} \emph{et~al.}
\newblock \bibinfo{title}{Hardware-efficient microwave-activated tunable
  coupling between superconducting qubits}.
\newblock \emph{\bibinfo{journal}{Phys. Rev. Lett.}}
  \textbf{\bibinfo{volume}{127}}, \bibinfo{pages}{200502}
  (\bibinfo{year}{2021}).

\bibitem{mkadzik2021precision}
\bibinfo{author}{M{\k{a}}dzik, M.~T.} \emph{et~al.}
\newblock \bibinfo{title}{Precision tomography of a three-qubit
  electron-nuclear quantum processor in silicon}.
\newblock \emph{\bibinfo{journal}{Nature}} \textbf{\bibinfo{volume}{601}},
  \bibinfo{pages}{348--353} (\bibinfo{year}{2022}).

\bibitem{wallman2016noise}
\bibinfo{author}{Wallman, J.~J.} \& \bibinfo{author}{Emerson, J.}
\newblock \bibinfo{title}{Noise tailoring for scalable quantum computation via
  randomized compiling}.
\newblock \emph{\bibinfo{journal}{Phys. Rev. A}} \textbf{\bibinfo{volume}{94}},
  \bibinfo{pages}{052325} (\bibinfo{year}{2016}).
\newblock \urlprefix\url{https://link.aps.org/doi/10.1103/PhysRevA.94.052325}.

\bibitem{ryan2021realization}
\bibinfo{author}{Ryan-Anderson, C.} \emph{et~al.}
\newblock \bibinfo{title}{Realization of real-time fault-tolerant quantum error
  correction}.
\newblock \emph{\bibinfo{journal}{Phys. Rev. X}} \textbf{\bibinfo{volume}{11}},
  \bibinfo{pages}{041058} (\bibinfo{year}{2021}).

\bibitem{chen2023transmon}
\bibinfo{author}{Chen, L.} \emph{et~al.}
\newblock \bibinfo{title}{Transmon qubit readout fidelity at the threshold for
  quantum error correction without a quantum-limited amplifier}.
\newblock \emph{\bibinfo{journal}{npj Quantum Inf.}}
  \textbf{\bibinfo{volume}{9}}, \bibinfo{pages}{26} (\bibinfo{year}{2023}).

\bibitem{hashim2021randomized}
\bibinfo{author}{Hashim, A.} \emph{et~al.}
\newblock \bibinfo{title}{Randomized compiling for scalable quantum computing
  on a noisy superconducting quantum processor}.
\newblock \emph{\bibinfo{journal}{Phys. Rev. X}} \textbf{\bibinfo{volume}{11}},
  \bibinfo{pages}{041039} (\bibinfo{year}{2021}).
\newblock \urlprefix\url{https://link.aps.org/doi/10.1103/PhysRevX.11.041039}.

\bibitem{erhard2019characterizing}
\bibinfo{author}{Erhard, A.} \emph{et~al.}
\newblock \bibinfo{title}{Characterizing large-scale quantum computers via
  cycle benchmarking}.
\newblock \emph{\bibinfo{journal}{Nat. Commun.}} \textbf{\bibinfo{volume}{10}},
  \bibinfo{pages}{1--7} (\bibinfo{year}{2019}).

\bibitem{hastings2016turning}
\bibinfo{author}{Hastings, M.~B.}
\newblock \bibinfo{title}{Turning gate synthesis errors into incoherent
  errors}.
\newblock \emph{\bibinfo{journal}{Preprint at
  https://arxiv.org/abs/1612.01011}}  (\bibinfo{year}{2016}).

\bibitem{campbell2017shorter}
\bibinfo{author}{Campbell, E.}
\newblock \bibinfo{title}{Shorter gate sequences for quantum computing by
  mixing unitaries}.
\newblock \emph{\bibinfo{journal}{Phys. Rev. A}} \textbf{\bibinfo{volume}{95}},
  \bibinfo{pages}{042306} (\bibinfo{year}{2017}).

\bibitem{cai2020mitigating}
\bibinfo{author}{Cai, Z.}, \bibinfo{author}{Xu, X.} \&
  \bibinfo{author}{Benjamin, S.~C.}
\newblock \bibinfo{title}{Mitigating coherent noise using pauli conjugation}.
\newblock \emph{\bibinfo{journal}{npj Quantum Inf.}}
  \textbf{\bibinfo{volume}{6}}, \bibinfo{pages}{1--9} (\bibinfo{year}{2020}).

\bibitem{ware2021experimental}
\bibinfo{author}{Ware, M.} \emph{et~al.}
\newblock \bibinfo{title}{Experimental pauli-frame randomization on a
  superconducting qubit}.
\newblock \emph{\bibinfo{journal}{Phys. Rev. A}}
  \textbf{\bibinfo{volume}{103}}, \bibinfo{pages}{042604}
  (\bibinfo{year}{2021}).

\bibitem{hashim2021optimized}
\bibinfo{author}{Hashim, A.} \emph{et~al.}
\newblock \bibinfo{title}{Optimized swap networks with equivalent circuit
  averaging for qaoa}.
\newblock \emph{\bibinfo{journal}{Phys. Rev. Res.}}
  \textbf{\bibinfo{volume}{4}}, \bibinfo{pages}{033028} (\bibinfo{year}{2022}).
\newblock
  \urlprefix\url{https://link.aps.org/doi/10.1103/PhysRevResearch.4.033028}.

\bibitem{chuang1997prescription}
\bibinfo{author}{Chuang, I.~L.} \& \bibinfo{author}{Nielsen, M.~A.}
\newblock \bibinfo{title}{Prescription for experimental determination of the
  dynamics of a quantum black box}.
\newblock \emph{\bibinfo{journal}{Journal of Modern Optics}}
  \textbf{\bibinfo{volume}{44}}, \bibinfo{pages}{2455--2467}
  (\bibinfo{year}{1997}).

\bibitem{nielsen2019python}
\bibinfo{author}{Nielsen, E.} \emph{et~al.}
\newblock \bibinfo{title}{Python gst implementation (pygsti) v. 0.9}.
\newblock \bibinfo{type}{Tech. Rep.}, \bibinfo{institution}{Sandia National
  Lab.(SNL-NM), Albuquerque, NM (United States)} (\bibinfo{year}{2019}).

\bibitem{nielsen2020probing}
\bibinfo{author}{Nielsen, E.} \emph{et~al.}
\newblock \bibinfo{title}{Probing quantum processor performance with pygsti}.
\newblock \emph{\bibinfo{journal}{Quantum Science and Technology}}
  \textbf{\bibinfo{volume}{5}}, \bibinfo{pages}{044002} (\bibinfo{year}{2020}).

\bibitem{viola1999dynamical}
\bibinfo{author}{Viola, L.}, \bibinfo{author}{Knill, E.} \&
  \bibinfo{author}{Lloyd, S.}
\newblock \bibinfo{title}{Dynamical decoupling of open quantum systems}.
\newblock \emph{\bibinfo{journal}{Phys. Rev. Lett.}}
  \textbf{\bibinfo{volume}{82}}, \bibinfo{pages}{2417} (\bibinfo{year}{1999}).

\bibitem{blume2022taxonomy}
\bibinfo{author}{Blume-Kohout, R.} \emph{et~al.}
\newblock \bibinfo{title}{A taxonomy of small markovian errors}.
\newblock \emph{\bibinfo{journal}{PRX Quantum}} \textbf{\bibinfo{volume}{3}},
  \bibinfo{pages}{020335} (\bibinfo{year}{2022}).

\bibitem{nielsen2021efficient}
\bibinfo{author}{Nielsen, E.}, \bibinfo{author}{Rudinger, K.},
  \bibinfo{author}{Proctor, T.}, \bibinfo{author}{Young, K.} \&
  \bibinfo{author}{Blume-Kohout, R.}
\newblock \bibinfo{title}{Efficient flexible characterization of quantum
  processors with nested error models}.
\newblock \emph{\bibinfo{journal}{New J. Phys.}} \textbf{\bibinfo{volume}{23}},
  \bibinfo{pages}{093020} (\bibinfo{year}{2021}).

\bibitem{rudinger2021experimental}
\bibinfo{author}{Rudinger, K.} \emph{et~al.}
\newblock \bibinfo{title}{Experimental characterization of crosstalk errors
  with simultaneous gate set tomography}.
\newblock \emph{\bibinfo{journal}{PRX Quantum}} \textbf{\bibinfo{volume}{2}},
  \bibinfo{pages}{040338} (\bibinfo{year}{2021}).
\newblock \urlprefix\url{https://link.aps.org/doi/10.1103/PRXQuantum.2.040338}.

\bibitem{wallman2015estimating}
\bibinfo{author}{Wallman, J.}, \bibinfo{author}{Granade, C.},
  \bibinfo{author}{Harper, R.} \& \bibinfo{author}{Flammia, S.~T.}
\newblock \bibinfo{title}{Estimating the coherence of noise}.
\newblock \emph{\bibinfo{journal}{New J. Phys.}} \textbf{\bibinfo{volume}{17}},
  \bibinfo{pages}{113020} (\bibinfo{year}{2015}).

\bibitem{boixo2018characterizing}
\bibinfo{author}{Boixo, S.} \emph{et~al.}
\newblock \bibinfo{title}{Characterizing quantum supremacy in near-term
  devices}.
\newblock \emph{\bibinfo{journal}{Nat. Phys.}} \textbf{\bibinfo{volume}{14}},
  \bibinfo{pages}{595--600} (\bibinfo{year}{2018}).

\bibitem{knill2004fault}
\bibinfo{author}{Knill, E.}
\newblock \bibinfo{title}{Fault-tolerant postselected quantum computation:
  Threshold analysis}.
\newblock \emph{\bibinfo{journal}{Preprint at quant-ph/0404104}}
  (\bibinfo{year}{2004}).

\bibitem{kern2005quantum}
\bibinfo{author}{Kern, O.}, \bibinfo{author}{Alber, G.} \&
  \bibinfo{author}{Shepelyansky, D.~L.}
\newblock \bibinfo{title}{Quantum error correction of coherent errors by
  randomization}.
\newblock \emph{\bibinfo{journal}{Eur. Phys. J. D}}
  \textbf{\bibinfo{volume}{32}}, \bibinfo{pages}{153--156}
  (\bibinfo{year}{2005}).

\bibitem{fowler2012surface}
\bibinfo{author}{Fowler, A.~G.}, \bibinfo{author}{Mariantoni, M.},
  \bibinfo{author}{Martinis, J.~M.} \& \bibinfo{author}{Cleland, A.~N.}
\newblock \bibinfo{title}{Surface codes: Towards practical large-scale quantum
  computation}.
\newblock \emph{\bibinfo{journal}{Phys. Rev. A}} \textbf{\bibinfo{volume}{86}},
  \bibinfo{pages}{032324} (\bibinfo{year}{2012}).

\bibitem{wang2003confinement}
\bibinfo{author}{Wang, C.}, \bibinfo{author}{Harrington, J.} \&
  \bibinfo{author}{Preskill, J.}
\newblock \bibinfo{title}{Confinement-higgs transition in a disordered gauge
  theory and the accuracy threshold for quantum memory}.
\newblock \emph{\bibinfo{journal}{Ann. Phys.}} \textbf{\bibinfo{volume}{303}},
  \bibinfo{pages}{31--58} (\bibinfo{year}{2003}).

\bibitem{raussendorf2006fault}
\bibinfo{author}{Raussendorf, R.}, \bibinfo{author}{Harrington, J.} \&
  \bibinfo{author}{Goyal, K.}
\newblock \bibinfo{title}{A fault-tolerant one-way quantum computer}.
\newblock \emph{\bibinfo{journal}{Ann. Phys.}} \textbf{\bibinfo{volume}{321}},
  \bibinfo{pages}{2242--2270} (\bibinfo{year}{2006}).

\bibitem{raussendorf2007topological}
\bibinfo{author}{Raussendorf, R.}, \bibinfo{author}{Harrington, J.} \&
  \bibinfo{author}{Goyal, K.}
\newblock \bibinfo{title}{Topological fault-tolerance in cluster state quantum
  computation}.
\newblock \emph{\bibinfo{journal}{New J. Phys.}} \textbf{\bibinfo{volume}{9}},
  \bibinfo{pages}{199} (\bibinfo{year}{2007}).

\bibitem{raussendorf2007fault}
\bibinfo{author}{Raussendorf, R.} \& \bibinfo{author}{Harrington, J.}
\newblock \bibinfo{title}{Fault-tolerant quantum computation with high
  threshold in two dimensions}.
\newblock \emph{\bibinfo{journal}{Phys. Rev. Lett.}}
  \textbf{\bibinfo{volume}{98}}, \bibinfo{pages}{190504}
  (\bibinfo{year}{2007}).

\bibitem{fowler2012towards}
\bibinfo{author}{Fowler, A.~G.}, \bibinfo{author}{Whiteside, A.~C.} \&
  \bibinfo{author}{Hollenberg, L.~C.}
\newblock \bibinfo{title}{Towards practical classical processing for the
  surface code}.
\newblock \emph{\bibinfo{journal}{Phys. Rev. Lett.}}
  \textbf{\bibinfo{volume}{108}}, \bibinfo{pages}{180501}
  (\bibinfo{year}{2012}).

\bibitem{lu2007quantum}
\bibinfo{author}{Lu, F.} \& \bibinfo{author}{Marinescu, D.~C.}
\newblock \bibinfo{title}{Quantum error correction of time-correlated errors}.
\newblock \emph{\bibinfo{journal}{Quantum Inf. Process.}}
  \textbf{\bibinfo{volume}{6}}, \bibinfo{pages}{273--293}
  (\bibinfo{year}{2007}).

\bibitem{wilen2021correlated}
\bibinfo{author}{Wilen, C.~D.} \emph{et~al.}
\newblock \bibinfo{title}{Correlated charge noise and relaxation errors in
  superconducting qubits}.
\newblock \emph{\bibinfo{journal}{Nature}} \textbf{\bibinfo{volume}{594}},
  \bibinfo{pages}{369--373} (\bibinfo{year}{2021}).

\bibitem{mundada2019suppression}
\bibinfo{author}{Mundada, P.}, \bibinfo{author}{Zhang, G.},
  \bibinfo{author}{Hazard, T.} \& \bibinfo{author}{Houck, A.}
\newblock \bibinfo{title}{Suppression of qubit crosstalk in a tunable coupling
  superconducting circuit}.
\newblock \emph{\bibinfo{journal}{Phys. Rev. App.}}
  \textbf{\bibinfo{volume}{12}}, \bibinfo{pages}{054023}
  (\bibinfo{year}{2019}).

\bibitem{zhao2020high}
\bibinfo{author}{Zhao, P.} \emph{et~al.}
\newblock \bibinfo{title}{High-contrast $zz$ interaction using superconducting
  qubits with opposite-sign anharmonicity}.
\newblock \emph{\bibinfo{journal}{Phys. Rev. Lett.}}
  \textbf{\bibinfo{volume}{125}}, \bibinfo{pages}{200503}
  (\bibinfo{year}{2020}).

\bibitem{ni2021scalable}
\bibinfo{author}{Ni, Z.} \emph{et~al.}
\newblock \bibinfo{title}{Scalable method for eliminating residual $zz$
  interaction between superconducting qubits}.
\newblock \emph{\bibinfo{journal}{Phys. Rev. Lett.}}
  \textbf{\bibinfo{volume}{129}}, \bibinfo{pages}{040502}
  (\bibinfo{year}{2022}).
\newblock
  \urlprefix\url{https://link.aps.org/doi/10.1103/PhysRevLett.129.040502}.

\bibitem{flammia2020efficient}
\bibinfo{author}{Flammia, S.~T.} \& \bibinfo{author}{Wallman, J.~J.}
\newblock \bibinfo{title}{Efficient estimation of pauli channels}.
\newblock \emph{\bibinfo{journal}{ACM Transactions on Quantum Computing}}
  \textbf{\bibinfo{volume}{1}}, \bibinfo{pages}{1--32} (\bibinfo{year}{2020}).

\bibitem{beale_stefanie_j_2020_3945250}
\bibinfo{author}{Beale, S.~J.} \emph{et~al.}
\newblock \bibinfo{title}{True-q} (\bibinfo{year}{2020}).
\newblock \urlprefix\url{https://doi.org/10.5281/zenodo.3945250}.

\bibitem{fowler2014quantifying}
\bibinfo{author}{Fowler, A.~G.} \& \bibinfo{author}{Martinis, J.~M.}
\newblock \bibinfo{title}{Quantifying the effects of local many-qubit errors
  and nonlocal two-qubit errors on the surface code}.
\newblock \emph{\bibinfo{journal}{Phys. Rev. A}} \textbf{\bibinfo{volume}{89}},
  \bibinfo{pages}{032316} (\bibinfo{year}{2014}).

\bibitem{ball2016effect}
\bibinfo{author}{Ball, H.}, \bibinfo{author}{Stace, T.~M.},
  \bibinfo{author}{Flammia, S.~T.} \& \bibinfo{author}{Biercuk, M.~J.}
\newblock \bibinfo{title}{Effect of noise correlations on randomized
  benchmarking}.
\newblock \emph{\bibinfo{journal}{Phys. Rev. A}} \textbf{\bibinfo{volume}{93}},
  \bibinfo{pages}{022303} (\bibinfo{year}{2016}).

\bibitem{ghosh2013understanding}
\bibinfo{author}{Ghosh, J.}, \bibinfo{author}{Fowler, A.~G.},
  \bibinfo{author}{Martinis, J.~M.} \& \bibinfo{author}{Geller, M.~R.}
\newblock \bibinfo{title}{Understanding the effects of leakage in
  superconducting quantum-error-detection circuits}.
\newblock \emph{\bibinfo{journal}{Phys. Rev. A}} \textbf{\bibinfo{volume}{88}},
  \bibinfo{pages}{062329} (\bibinfo{year}{2013}).

\bibitem{wallman2016robust}
\bibinfo{author}{Wallman, J.~J.}, \bibinfo{author}{Barnhill, M.} \&
  \bibinfo{author}{Emerson, J.}
\newblock \bibinfo{title}{Robust characterization of leakage errors}.
\newblock \emph{\bibinfo{journal}{New J. Phys.}} \textbf{\bibinfo{volume}{18}},
  \bibinfo{pages}{043021} (\bibinfo{year}{2016}).

\bibitem{chen2016measuring}
\bibinfo{author}{Chen, Z.} \emph{et~al.}
\newblock \bibinfo{title}{Measuring and suppressing quantum state leakage in a
  superconducting qubit}.
\newblock \emph{\bibinfo{journal}{Phys. Rev. Lett.}}
  \textbf{\bibinfo{volume}{116}}, \bibinfo{pages}{020501}
  (\bibinfo{year}{2016}).

\bibitem{wood2018quantification}
\bibinfo{author}{Wood, C.~J.} \& \bibinfo{author}{Gambetta, J.~M.}
\newblock \bibinfo{title}{Quantification and characterization of leakage
  errors}.
\newblock \emph{\bibinfo{journal}{Phys. Rev. A}} \textbf{\bibinfo{volume}{97}},
  \bibinfo{pages}{032306} (\bibinfo{year}{2018}).

\bibitem{hayes2020eliminating}
\bibinfo{author}{Hayes, D.} \emph{et~al.}
\newblock \bibinfo{title}{Eliminating leakage errors in hyperfine qubits}.
\newblock \emph{\bibinfo{journal}{Phys. Rev. Lett.}}
  \textbf{\bibinfo{volume}{124}}, \bibinfo{pages}{170501}
  (\bibinfo{year}{2020}).

\bibitem{babu2021state}
\bibinfo{author}{Babu, A.~P.}, \bibinfo{author}{Tuorila, J.} \&
  \bibinfo{author}{Ala-Nissila, T.}
\newblock \bibinfo{title}{State leakage during fast decay and control of a
  superconducting transmon qubit}.
\newblock \emph{\bibinfo{journal}{npj Quantum Inf.}}
  \textbf{\bibinfo{volume}{7}}, \bibinfo{pages}{1--8} (\bibinfo{year}{2021}).

\bibitem{proctor2020detecting}
\bibinfo{author}{Proctor, T.} \emph{et~al.}
\newblock \bibinfo{title}{Detecting and tracking drift in quantum information
  processors}.
\newblock \emph{\bibinfo{journal}{Nat. Commun.}} \textbf{\bibinfo{volume}{11}},
  \bibinfo{pages}{1--9} (\bibinfo{year}{2020}).

\bibitem{serniak2018hot}
\bibinfo{author}{Serniak, K.} \emph{et~al.}
\newblock \bibinfo{title}{Hot nonequilibrium quasiparticles in transmon
  qubits}.
\newblock \emph{\bibinfo{journal}{Phys. Rev. Lett.}}
  \textbf{\bibinfo{volume}{121}}, \bibinfo{pages}{157701}
  (\bibinfo{year}{2018}).

\bibitem{de2020two}
\bibinfo{author}{de~Graaf, S.} \emph{et~al.}
\newblock \bibinfo{title}{Two-level systems in superconducting quantum devices
  due to trapped quasiparticles}.
\newblock \emph{\bibinfo{journal}{Sci. Adv.}} \textbf{\bibinfo{volume}{6}},
  \bibinfo{pages}{eabc5055} (\bibinfo{year}{2020}).

\bibitem{berlin2022changes}
\bibinfo{author}{Berlin-Udi, M.} \emph{et~al.}
\newblock \bibinfo{title}{Changes in electric field noise due to thermal
  transformation of a surface ion trap}.
\newblock \emph{\bibinfo{journal}{Phys. Rev. B}}
  \textbf{\bibinfo{volume}{106}}, \bibinfo{pages}{035409}
  (\bibinfo{year}{2022}).

\bibitem{webb2018resilient}
\bibinfo{author}{Webb, A.~E.} \emph{et~al.}
\newblock \bibinfo{title}{Resilient entangling gates for trapped ions}.
\newblock \emph{\bibinfo{journal}{Phys. Rev. Lett.}}
  \textbf{\bibinfo{volume}{121}}, \bibinfo{pages}{180501}
  (\bibinfo{year}{2018}).

\bibitem{blume2020wildcard}
\bibinfo{author}{Blume-Kohout, R.}, \bibinfo{author}{Rudinger, K.},
  \bibinfo{author}{Nielsen, E.}, \bibinfo{author}{Proctor, T.} \&
  \bibinfo{author}{Young, K.}
\newblock \bibinfo{title}{Wildcard error: Quantifying unmodeled errors in
  quantum processors}.
\newblock \emph{\bibinfo{journal}{Preprint at
  https://arxiv.org/abs/2012.12231}}  (\bibinfo{year}{2020}).

\bibitem{wootton2011bringing}
\bibinfo{author}{Wootton, J.~R.} \& \bibinfo{author}{Pachos, J.~K.}
\newblock \bibinfo{title}{Bringing order through disorder: Localization of
  errors in topological quantum memories}.
\newblock \emph{\bibinfo{journal}{Phys. Rev. Lett.}}
  \textbf{\bibinfo{volume}{107}}, \bibinfo{pages}{030503}
  (\bibinfo{year}{2011}).

\bibitem{stark2011localization}
\bibinfo{author}{Stark, C.}, \bibinfo{author}{Pollet, L.},
  \bibinfo{author}{Imamo{\u{g}}lu, A.} \& \bibinfo{author}{Renner, R.}
\newblock \bibinfo{title}{Localization of toric code defects}.
\newblock \emph{\bibinfo{journal}{Phys. Rev. Lett.}}
  \textbf{\bibinfo{volume}{107}}, \bibinfo{pages}{030504}
  (\bibinfo{year}{2011}).

\bibitem{bravyi2012disorder}
\bibinfo{author}{Bravyi, S.} \& \bibinfo{author}{K{\"o}nig, R.}
\newblock \bibinfo{title}{Disorder-assisted error correction in majorana
  chains}.
\newblock \emph{\bibinfo{journal}{Commun. Math. Phys.}}
  \textbf{\bibinfo{volume}{316}}, \bibinfo{pages}{641--692}
  (\bibinfo{year}{2012}).

\bibitem{bravyi2018correcting}
\bibinfo{author}{Bravyi, S.}, \bibinfo{author}{Englbrecht, M.},
  \bibinfo{author}{K{\"o}nig, R.} \& \bibinfo{author}{Peard, N.}
\newblock \bibinfo{title}{Correcting coherent errors with surface codes}.
\newblock \emph{\bibinfo{journal}{npj Quantum Inf.}}
  \textbf{\bibinfo{volume}{4}}, \bibinfo{pages}{1--6} (\bibinfo{year}{2018}).

\bibitem{geller2013efficient}
\bibinfo{author}{Geller, M.~R.} \& \bibinfo{author}{Zhou, Z.}
\newblock \bibinfo{title}{Efficient error models for fault-tolerant
  architectures and the pauli twirling approximation}.
\newblock \emph{\bibinfo{journal}{Phys. Rev. A}} \textbf{\bibinfo{volume}{88}},
  \bibinfo{pages}{012314} (\bibinfo{year}{2013}).

\bibitem{tomita2014low}
\bibinfo{author}{Tomita, Y.} \& \bibinfo{author}{Svore, K.~M.}
\newblock \bibinfo{title}{Low-distance surface codes under realistic quantum
  noise}.
\newblock \emph{\bibinfo{journal}{Phys. Rev. A}} \textbf{\bibinfo{volume}{90}},
  \bibinfo{pages}{062320} (\bibinfo{year}{2014}).

\bibitem{katabarwa2015logical}
\bibinfo{author}{Katabarwa, A.} \& \bibinfo{author}{Geller, M.~R.}
\newblock \bibinfo{title}{Logical error rate in the pauli twirling
  approximation}.
\newblock \emph{\bibinfo{journal}{Sci. Rep.}} \textbf{\bibinfo{volume}{5}},
  \bibinfo{pages}{1--6} (\bibinfo{year}{2015}).

\bibitem{puzzuoli2014tractable}
\bibinfo{author}{Puzzuoli, D.} \emph{et~al.}
\newblock \bibinfo{title}{Tractable simulation of error correction with honest
  approximations to realistic fault models}.
\newblock \emph{\bibinfo{journal}{Phys. Rev. A}} \textbf{\bibinfo{volume}{89}},
  \bibinfo{pages}{022306} (\bibinfo{year}{2014}).

\bibitem{gutierrez2015comparison}
\bibinfo{author}{Guti{\'e}rrez, M.} \& \bibinfo{author}{Brown, K.~R.}
\newblock \bibinfo{title}{Comparison of a quantum error-correction threshold
  for exact and approximate errors}.
\newblock \emph{\bibinfo{journal}{Phys. Rev. A}} \textbf{\bibinfo{volume}{91}},
  \bibinfo{pages}{022335} (\bibinfo{year}{2015}).

\bibitem{katabarwa2017dynamical}
\bibinfo{author}{Katabarwa, A.}
\newblock \bibinfo{title}{A dynamical interpretation of the pauli twirling
  approximation and quantum error correction}.
\newblock \emph{\bibinfo{journal}{Preprint at
  https://arxiv.org/abs/1701.03708}}  (\bibinfo{year}{2017}).

\bibitem{fern2006generalized}
\bibinfo{author}{Fern, J.}, \bibinfo{author}{Kempe, J.},
  \bibinfo{author}{Simic, S.~N.} \& \bibinfo{author}{Sastry, S.}
\newblock \bibinfo{title}{Generalized performance of concatenated quantum
  codes—a dynamical systems approach}.
\newblock \emph{\bibinfo{journal}{IEEE Trans. Autom. Control}}
  \textbf{\bibinfo{volume}{51}}, \bibinfo{pages}{448--459}
  (\bibinfo{year}{2006}).

\bibitem{gutierrez2016errors}
\bibinfo{author}{Guti{\'e}rrez, M.}, \bibinfo{author}{Smith, C.},
  \bibinfo{author}{Lulushi, L.}, \bibinfo{author}{Janardan, S.} \&
  \bibinfo{author}{Brown, K.~R.}
\newblock \bibinfo{title}{Errors and pseudothresholds for incoherent and
  coherent noise}.
\newblock \emph{\bibinfo{journal}{Phys. Rev. A}} \textbf{\bibinfo{volume}{94}},
  \bibinfo{pages}{042338} (\bibinfo{year}{2016}).

\bibitem{darmawan2017tensor}
\bibinfo{author}{Darmawan, A.~S.} \& \bibinfo{author}{Poulin, D.}
\newblock \bibinfo{title}{Tensor-network simulations of the surface code under
  realistic noise}.
\newblock \emph{\bibinfo{journal}{Phys. Rev. Lett.}}
  \textbf{\bibinfo{volume}{119}}, \bibinfo{pages}{040502}
  (\bibinfo{year}{2017}).

\bibitem{greenbaum2017modeling}
\bibinfo{author}{Greenbaum, D.} \& \bibinfo{author}{Dutton, Z.}
\newblock \bibinfo{title}{Modeling coherent errors in quantum error
  correction}.
\newblock \emph{\bibinfo{journal}{Quantum Science and Technology}}
  \textbf{\bibinfo{volume}{3}}, \bibinfo{pages}{015007} (\bibinfo{year}{2017}).

\bibitem{hakkaku2021sampling}
\bibinfo{author}{Hakkaku, S.}, \bibinfo{author}{Mitarai, K.} \&
  \bibinfo{author}{Fujii, K.}
\newblock \bibinfo{title}{Sampling-based quasiprobability simulation for
  fault-tolerant quantum error correction on the surface codes under coherent
  noise}.
\newblock \emph{\bibinfo{journal}{Phys. Rev. Res.}}
  \textbf{\bibinfo{volume}{3}}, \bibinfo{pages}{043130} (\bibinfo{year}{2021}).

\bibitem{huang2019performance}
\bibinfo{author}{Huang, E.}, \bibinfo{author}{Doherty, A.~C.} \&
  \bibinfo{author}{Flammia, S.}
\newblock \bibinfo{title}{Performance of quantum error correction with coherent
  errors}.
\newblock \emph{\bibinfo{journal}{Phys. Rev. A}} \textbf{\bibinfo{volume}{99}},
  \bibinfo{pages}{022313} (\bibinfo{year}{2019}).

\bibitem{dehollain2016optimization}
\bibinfo{author}{Dehollain, J.~P.} \emph{et~al.}
\newblock \bibinfo{title}{Optimization of a solid-state electron spin qubit
  using gate set tomography}.
\newblock \emph{\bibinfo{journal}{New J. Phys.}} \textbf{\bibinfo{volume}{18}},
  \bibinfo{pages}{103018} (\bibinfo{year}{2016}).

\bibitem{hughes2020benchmarking}
\bibinfo{author}{Hughes, A.} \emph{et~al.}
\newblock \bibinfo{title}{Benchmarking a high-fidelity mixed-species entangling
  gate}.
\newblock \emph{\bibinfo{journal}{Phys. Rev. Lett.}}
  \textbf{\bibinfo{volume}{125}}, \bibinfo{pages}{080504}
  (\bibinfo{year}{2020}).

\bibitem{white2021performance}
\bibinfo{author}{White, G.~A.}, \bibinfo{author}{Hill, C.~D.} \&
  \bibinfo{author}{Hollenberg, L.~C.}
\newblock \bibinfo{title}{Performance optimization for drift-robust fidelity
  improvement of two-qubit gates}.
\newblock \emph{\bibinfo{journal}{Phys. Rev. App.}}
  \textbf{\bibinfo{volume}{15}}, \bibinfo{pages}{014023}
  (\bibinfo{year}{2021}).

\bibitem{goss2023extending}
\bibinfo{author}{Goss, N.} \emph{et~al.}
\newblock \bibinfo{title}{Extending the computational reach of a
  superconducting qutrit processor}.
\newblock \emph{\bibinfo{journal}{Preprint at
  https://arxiv.org/abs/2305.16507}}  (\bibinfo{year}{2023}).

\bibitem{ferracin2022efficiently}
\bibinfo{author}{Ferracin, S.} \emph{et~al.}
\newblock \bibinfo{title}{Efficiently improving the performance of noisy
  quantum computers}.
\newblock \emph{\bibinfo{journal}{Preprint at
  https://arxiv.org/abs/2201.10672}}  (\bibinfo{year}{2022}).

\bibitem{wilks1938large}
\bibinfo{author}{Wilks, S.~S.}
\newblock \bibinfo{title}{The large-sample distribution of the likelihood ratio
  for testing composite hypotheses}.
\newblock \emph{\bibinfo{journal}{Ann. Math. Stat.}}
  \textbf{\bibinfo{volume}{9}}, \bibinfo{pages}{60--62} (\bibinfo{year}{1938}).

\bibitem{proctor2021measuring}
\bibinfo{author}{Proctor, T.}, \bibinfo{author}{Rudinger, K.},
  \bibinfo{author}{Young, K.}, \bibinfo{author}{Nielsen, E.} \&
  \bibinfo{author}{Blume-Kohout, R.}
\newblock \bibinfo{title}{Measuring the capabilities of quantum computers}.
\newblock \emph{\bibinfo{journal}{Nat. Phys.}} \bibinfo{pages}{1--5}
  (\bibinfo{year}{2021}).

\end{thebibliography}


\end{document}